\documentclass[preprint,sort&compress,3p,times]{elsarticle}
\usepackage{caption}
\usepackage[caption=false]{subfig}
\usepackage{amsmath, amsthm, amssymb}
\usepackage{mathtools}
\usepackage{graphicx}
\usepackage{float}
\usepackage{breqn}
\usepackage{bm}
\usepackage{url}
\newcommand\numberthis{\addtocounter{equation}{1}\tag{\theequation}}

\numberwithin{equation}{section}

\begin{document}

\begin{frontmatter}
\title{Complex-Plane Generalization of Scalar Levin Transforms: A Robust, Rapidly Convergent Method to Compute Potentials and Fields in Multi-Layered Media}

\author[ad1]{Kamalesh Sainath\corref{cor1}}
\ead{sainath.1@osu.edu}

\author[ad1]{Fernando L. Teixeira}
\ead{teixeira@ece.osu.edu}

\author[ad2]{Burkay Donderici}
\ead{Burkay.Donderici@Halliburton.com}

\cortext[cor1]{Corresponding author}

\address[ad1]{The Ohio State University, Electroscience Laboratory, 1330 Kinnear Road, Columbus, Ohio, USA 43212}

\address[ad2]{Halliburton, Sensor Physics and Technology, 3000 N. Sam Houston Pkwy E, Houston, TX, USA 77032}

\begin{abstract}

\noindent
We propose the complex-plane generalization of a powerful algebraic sequence acceleration algorithm, the Method of Weighted Averages (MWA), to guarantee \emph{exponential-cum-algebraic} convergence of Fourier and Fourier-Hankel (F-H) integral transforms. This ``complex-plane" MWA, effected via a linear-path detour in the complex plane, results in rapid, absolute convergence of field/potential solutions in multi-layered environments \emph{regardless} of the source-observer geometry and anisotropy/loss of the media present. In this work, we first introduce a new integration path used to evaluate the field contribution arising from the radiation spectra. Subsequently, we (1) exhibit the foundational relations behind the complex-plane extension to a general Levin-type sequence convergence accelerator, (2) specialize this analysis to one member of the Levin transform family (the MWA), (3) address and circumvent restrictions, arising for two-dimensional integrals associated with wave dynamics problems, through minimal complex-plane detour restrictions and a novel partition of the integration domain, (4) develop and compare two formulations based on standard/real-axis MWA variants, and (5) present validation results and convergence characteristics for one of these two formulations.
\end{abstract}


\end{frontmatter}

\section{\label{intro}Introduction}
In many application areas concerning time-harmonic electromagnetic (EM) fields, one encounters environments containing media of varying and arbitrary anisotropy\footnote{We assume each medium's anisotropy manifests in \emph{diagonalizable} constitutive material tensors to ensure completeness of the plane wave basis. Since all naturally-occurring media possess diagonalizable material tensors, in practical applications this assumption is always true.} whose inhomogeneity can be approximated as multi-layered in nature. Examples include geophysical prospection \cite{sofia,wei,anderson1,howard,zhdanov,wang,moran1}, plasma physics \cite{paulus}, antenna design \cite{pozar,pozar2}, optical field control \cite{jain}, microwave remote sensing \cite{jehle}, ground-penetrating radar ~\cite{lambot1,lambot2}, and microwave circuits \cite{mosig2}, among others. Such applications regularly encounter integrals of the form\footnote{Appendix A summarizes the notation, terminology, and conventions used here.}
 \begin{equation}\label{SI1} f(\bold{r}) \ \sim \iint\limits_{-\infty}^{+\infty} \tilde{f}(k_x,k_y)\mathrm{e}^{ik_x(x-x')+ik_y(y-y')+i\tilde{k}_z(z-z')}\,\mathrm{d} \, k_x \,\mathrm{d} \, k_y \end{equation}
and/or
 \begin{equation}\label{SI2} f(\bold{r}) \ \sim \int\limits_{-\infty}^{+\infty} \tilde{f}(k_{\rho})H^{(1)}_n(k_{\rho}|\rho-\rho'|) \mathrm{e}^{i\tilde{k}_z(z-z')}\,\mathrm{d} \, k_{\rho} \end{equation}
which express space-domain field/potential functions as Fourier and Fourier-Hankel (F-H) integral transforms (resp.).

In many practical applications, these integrals must often be rapidly evaluated for a wide range of longitudinal and transverse source-observer separation geometries $\bold{r}-\bold{r}' \neq \bold{0}$ (e.g. for potential or field profile reconstruction). However, when using standard integration paths that run on/close to the real axis such as (1) the classic Sommerfeld Integration Path (SIP) \cite{chewch2} and (2) paths detouring around the branch points, branch cuts, and poles followed by real-axis integration \cite{mich1,mich2,sainath}, the convergence rate of these integrals is strongly dependent upon the transverse ($\bold{r}_t-\bold{r}_t'$) and longitudinal ($z-z'$) separations. $\bold{r}_t-\bold{r}_t'$ determines the rapidity of the integrand's oscillation due to the Fourier and/or Hankel kernels in \eqref{SI1}-\eqref{SI2}, with rising $| \rho-\rho' |$ leading to an integrand that traditionally requires increasingly finer sampling to limit spatial aliasing and thus leads to undesirably long computation times. Furthermore, the longitudinal separation $z-z'$ governs the rate at which the evanescent spectrum's field contribution decays with increasing transverse wave number magnitudes\footnote{i.e. $|k_x|$ and $|k_y|$ for Fourier double-integrals, or $|k_{\rho}|$ for F-H integrals.}, with rising $|z-z'|$ effecting more rapid decay (and hence faster convergence) \cite{mosig1}. On the other hand, as $|z-z'| \to 0$ the convergence rate lessens, with the limiting case $z-z'=0$ yielding integrals of the form
 \begin{equation}\label{SI3} f(\bold{r}) \ \sim \iint\limits_{-\infty}^{+\infty} \tilde{f}(k_x,k_y)\mathrm{e}^{ik_x(x-x')+ik_y(y-y')}\,\mathrm{d} \, k_x \,\mathrm{d} \, k_y \end{equation}
and
 \begin{equation}\label{SI4} f(\bold{r}) \ \sim \int\limits_{-\infty}^{+\infty} \tilde{f}(k_{\rho})H^{(1)}_n(k_{\rho}|\rho-\rho'|) \,\mathrm{d} \, k_{\rho} \end{equation}
that lead to divergent results when numerically evaluated, using these standard paths, without convergence acceleration.

\begin{figure}[H]
\centering
\subfloat[\label{Well1}]{\includegraphics[width=3in]{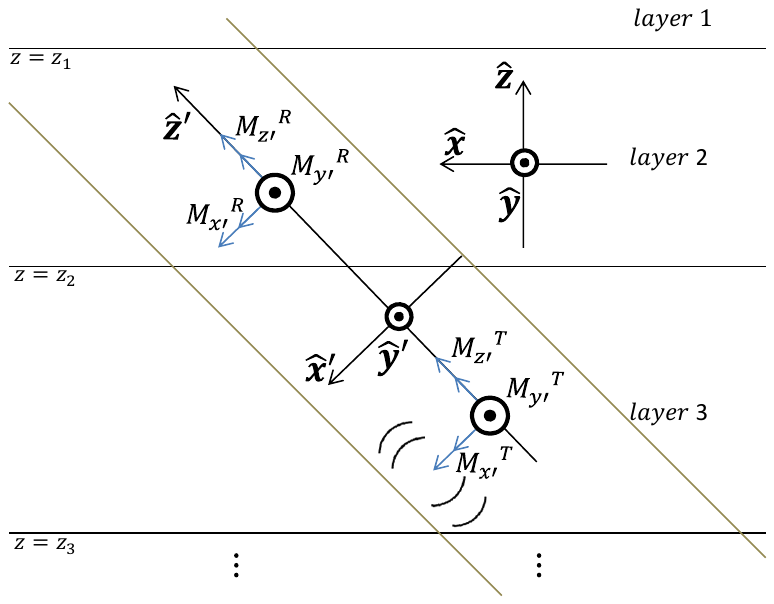}}

\subfloat[\label{Well2}]{\includegraphics[width=3in]{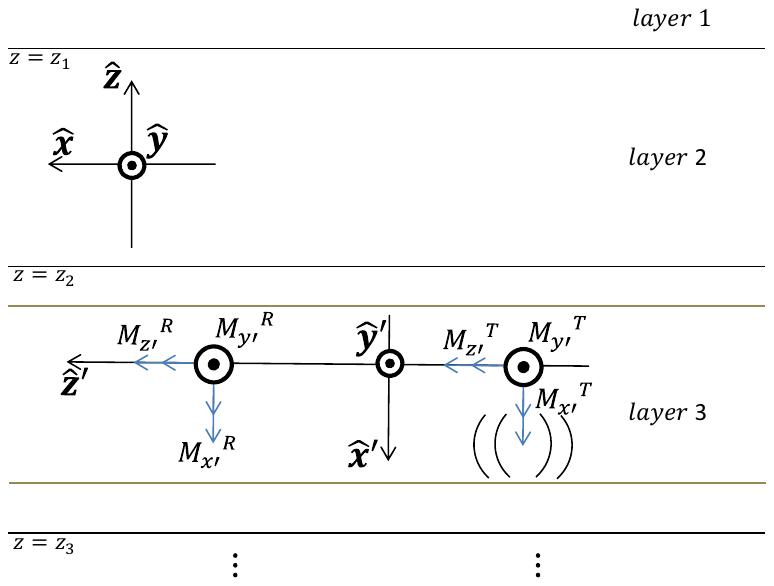}}

\subfloat[\label{MSubstrate}]{\includegraphics[width=3in]{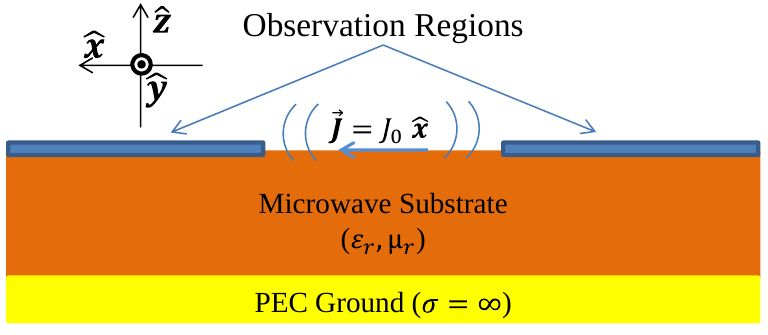}}
\caption{\small Figure \ref{Well1} depicts a ``triaxial" hydrocarbon sensor system \cite{zhdanov} of three loop antenna transmitters $\{M^T \}$ and three loop antenna receivers $\{M^R \}$ traversing a vertical/moderately-inclined logging path bounded by a borehole (dark gold lines). Here, one typically finds $|z-z'|$ large enough to use standard numerical integration methods, based on real-/near real-axis paths, without convergence acceleration. On the other hand, Figure \ref{Well2} shows the same sensor system traversing a horizontal path while Figure \ref{MSubstrate} exhibits a micro-strip geometry in which the user requests the field distribution at the air-substrate interface. The two latter geometries exhibit $0 \leq |z-z'| \ll 1$ and represent scenarios for which these standard methods typically yield divergent results due to the oscillatory-divergent nature of integrals like \eqref{SI3}-\eqref{SI4}.}
\label{AppFigs}
    \end{figure}

See Figure \ref{AppFigs} for typical application scenarios wherein these standard paths either succeed or fail to deliver accurate field results. Observing Figure \ref{AppFigs}, one immediately realizes that devising an evaluation method for these integrals exhibiting \emph{robustness} with respect to all ranges of $\bold{r}-\bold{r}' \neq \bold{0}$ and medium classes (e.g. isotropic, uniaxial, biaxial) is highly desirable. This robustness criterion inherently excludes fundamentally approximate methods such as image and asymptotic methods due to their geometry-specific applicability and lack of rigorous error control \cite{mosig7,chewch2,mich1,mich5,caboussat}. As a result, to reliably ensure accurate field results for arbitrary environmental medium composition/source-observer geometry combinations, we choose a direct numerical integration method.

In this vein, one option involves pairing standard integration methods with (real-axis path based) algebraic convergence acceleration techniques such as the standard MWA which, based on published numerical results, successfully imparts algebraic convergence acceleration even when $|z-z'|=0$ \cite{mosig1,mich2}. However, it is desirable to (1) guarantee absolute, exponential convergence in the classical/Riemann sense for \emph{any} $\bold{r}-\bold{r}' \neq \bold{0}$ separation geometry (in contrast to only guaranteeing algebraic convergence in the Abel sense when $|z-z'|=0$ \cite{mosig1}) and (2) endow error control to the evanescent-zone field contribution associated with the tail integral, whose relative importance (compared to the radiation-zone contribution) to the field solution grows as $|\bold{r}-\bold{r}'|$ decreases, to ensure that both the radiation-zone and evanescent-zone contributions are accurately evaluated\footnote{One cannot rely upon a-posteriori error checking, as was done in \cite{mich2,mosig1}, for general environment/source-observer scenarios.}. To this end, we propose a novel numerical integration method, representing a complex-plane generalization of a specific member of the ``scalar Levin-type sequence transform" (SLST) family \cite{homeier} (i.e. the MWA), that:
\begin{enumerate}
\item bends the ``extrapolation region"/tail \cite{sainath} integration path sections to guarantee absolute, exponential convergence of integrals like \eqref{SI1}-\eqref{SI4},
\item imparts added, \emph{robust} algebraic convergence acceleration to the tail integrals, which compounds with the exponential convergence acceleration to effect absolute, \emph{exponential-cum-algebraic} convergence, via use of a linear path bend combined with our novel, complex-plane generalization of the MWA \cite{mich2,mosig1},
\item adjusts the detour bend angles to account for the presence of branch points, branch cuts, and poles (summarily referred to here as ``critical points"), and
\item addresses the added challenges associated with evaluating \emph{two-dimensional} integral transforms arising as solutions to the wave equation in planar-stratified environments lacking azimuthal symmetry.
\end{enumerate}
 We note that other path deformation techniques, such as the Steepest Descent Path (SDP) and one comprising the enclosure of the first/fourth quadrants of the $k_{\rho}$ plane involving an imaginary-axis integration, have been investigated and used \cite{mosig5,mosig6,chew2,chewch2}. However, we seek a \emph{robust} integration method, valid for all $\bold{r}-\bold{r}' \neq \bold{0}$ geometries, that obviates having to separately account for discrete poles while possessing applicability to multi-layered environments containing media with arbitrary anisotropy and loss. Thus while our method may result in longer solution times versus above-mentioned methods, it touts general applicability and minimal necessary book-keeping as its defining virtues.

 Furthermore, a robust detour path within the pre-extrapolation region \cite{sainath} maintaining a near-constant separation between the path and critical points near/on the real axis would be preferred over more traditional paths used with the MWA \cite{mich1,sainath}. To address this, the paper's second contribution entails a trapezoidal integration path paired with adaptive $hp$ refinement\footnote{The adaptive $hp$ refinement integration methodology is the same as in \cite{sainath}, and thus is not discussed further here.}.

  In Section \ref{preext} we present and discuss our revision to the radiation-zone integration path\footnote{We present this secondary contribution first for fluidity in the narrative.}. In Sections \ref{ext} and \ref{weightcomp} we develop the detoured linear integration path and complex-plane generalization to SLST for efficiently evaluating the tail sections of \eqref{SI1}-\eqref{SI4}, as well as exhibit and compare two possible candidate formulations to implement the resulting modified-MWA. These developments are formulated in the context of two-dimensional integrals such as \eqref{SI1} to simultaneously address herein their additional issues versus one-dimensional integrals. However, the formulation applies equally to one-dimensional F-H transforms like \eqref{SI2} appearing in field/potential computations within cylindrically- and (azimuthal-symmetric) planar-stratified environments and, after converting the Fourier-Bessel (F-B) transform to a F-H transform \cite{chewch2,chewch3,moon}, to F-B transforms as well\footnote{One can accommodate the logarithmic branch-cut, manifest on the -Re[$k_{\rho}$] axis for F-H transforms \cite{chewch2}, through a slight perturbation of the Re$[k_{\rho}]<0$ half-plane path into the second quadrant.}. Section \ref{results} presents validation results using one of the two new formulations. In Section \ref{conv} we present a study on the convergence characteristics of our algorithm as concerning the same formulation used to generate the results in Section \ref{results}. Finally, Section \ref{conclusion} contains our concluding remarks.

 In the ensuing discussion, we assume appropriate transformations to the material tensors and source vector have already been performed to effect a coordinate rotation such that in the resultant (azimuthal-rotated) coordinate frame, within which all integration is performed, one has $x-x'=y-y' \geq 0$ \footnote{More generally, if $|x-x'|=|y-y'|$ in the rotated frame the method will work. Of course, rotating such that $x-x'\leq0$ forces one to alter the $k_x$ plane extrapolation region path such that it now incurs into the Im[$k_x]<0$ half-plane (and similarly for $k_y$, $y-y'$).}. Discussed in detail at the end of Section \ref{ext}, this is done to guarantee absolute convergence and maximize exponential decay of both the $k_x$ \emph{and} $k_y$ integrals.
\section{\label{preext}Pre-Extrapolation Region Path Revision}
\begin{figure}[H]
\centering
\subfloat[\label{CFig1}]{\includegraphics[width=3in]{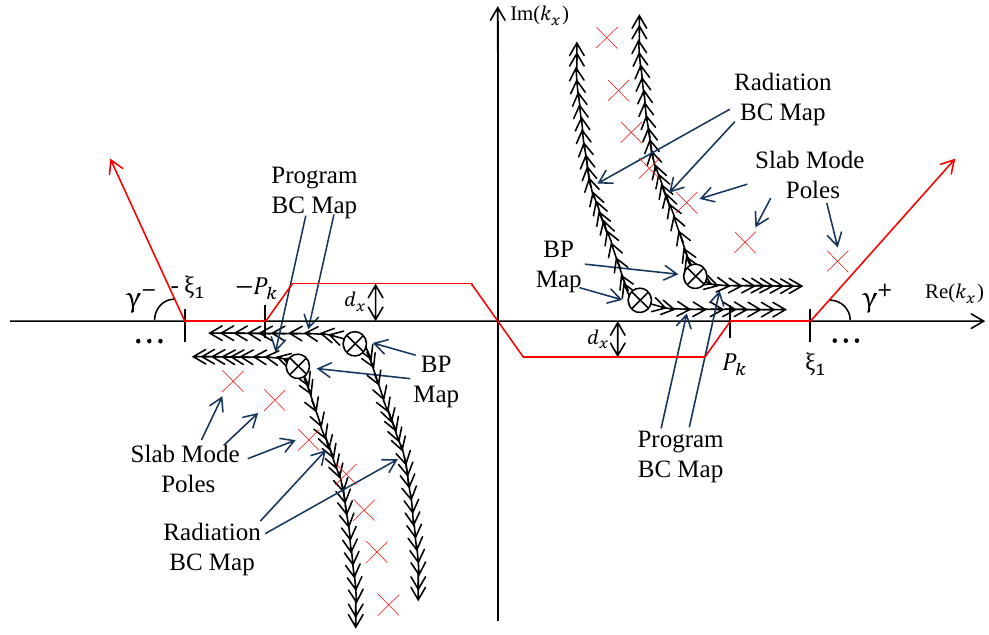}}

\subfloat[\label{CFig2}]{\includegraphics[width=3in]{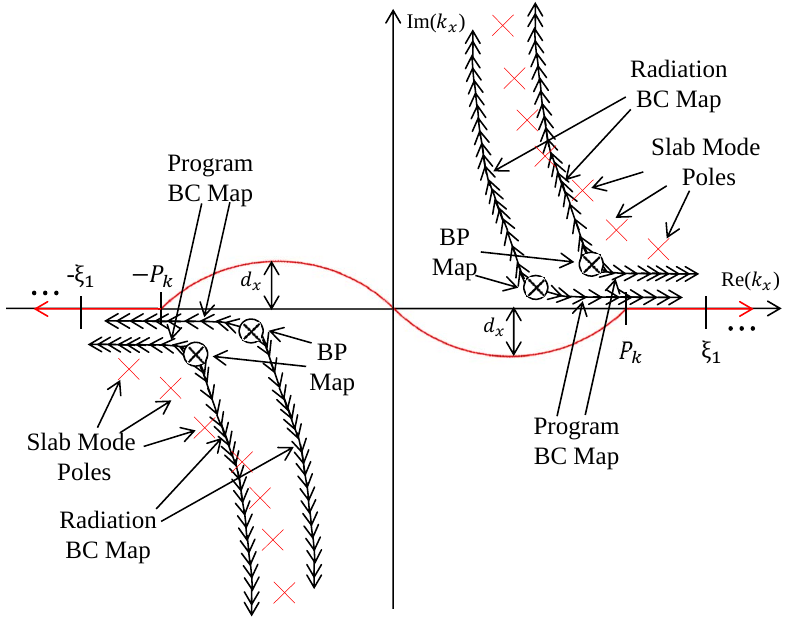}}
\caption{\small Figures \ref{CFig1} and \ref{CFig2} depict the new and old integration $k_x$ plane integration paths used in this paper and \cite{sainath} (resp.). ``Radiation BC Map" and ``Program BC Map" refer to the branch cuts associated with the radiation/boundedness condition at infinity (Im[$\tilde{k}_z^2]=0$, Re[$\tilde{k}_z^2]>0$) and the computer program's square root convention (Im[$\tilde{k}_z^2]=0$, Re[$\tilde{k}_z^2]<0$) (resp.). The encircled ``X" symbols represent branch points and the red ``X" symbols represent guided mode poles. For $K$ extrapolation intervals used in the bottom or top method, the red contour represents the integration path connecting the end-points $k_x=(-\xi_1 - K\Delta \xi_x,\xi_1 + K\Delta \xi_x)$ or $k_x=(-\xi_1 - t_o^-K\Delta \xi^{-'},\xi_1 + t_o^+K\Delta \xi^{+'})$ (resp.); see Sections \ref{preext}-\ref{ext} for definitions of $\Delta \xi_x$, $\Delta \xi_x^{+'}$, $\Delta \xi_x^{-'}$, $t_o^+$, and $t_o^-$.}
\label{ContourFigs}
    \end{figure}

First we discuss the parameterization and initial sub-division of the $k_x$ plane pre-extrapolation region; discussion of the $k_y$ plane follows identically due to our assuming $x-x'=y-y' \geq 0$. Applying a parameterization similar to that in \cite{sainath,mich1}, define $\pm {P_{k}}$ as the points on the Re$[k_x$] axis within which one detours, $d_{x}$ as the maximum height of the trapezoid-shaped detour, and $\pm \xi_1$ as the points on the Re$[k_x$] axis within which one adaptively integrates (see Figure \ref{ContourFigs}).

 To compute $\pm P_k$, first define $n^+$ as the magnitude of the real part of the global ``effective" refractive index among all the layers (see \cite{sainath} on computing $n^+$). One then computes $P_k$ analogously to \cite{mich1} and sets $P_k=l_ok_o(n^++1)$, where $l_o \geq 1$ is a user-defined pre-extrapolation region magnification constant\footnote{\label{ben1}This detour allows magnification of $P_k$ without compromising the detour height near critical points, which represents one of two primary benefits compared to the half-sine-shaped contour \cite{mich1,sainath}.}. Next, define $\Delta x=|x-x'|=\Delta y=|y-y'|$, $\Delta z=|z-z'|$, $a'$, and $b'$, where $0<a'<1/2$ and $b'>1$. Now compute the following pre-extrapolation region integration path parameters \cite{mich1,sainath}\footnote{$T_0>0$ limits the extrapolation region sub-interval length $\Delta \xi_x$ when $\Delta x \ll 1$ to ensure the extrapolation intervals (see Sections \ref{preext}-\ref{ext}) are adequately sampled, thereby limiting spatial aliasing.}:
\begin{align}  Q_k &= a'k_o(n^++1) \numberthis \label{ReDx} \\
 d_x &=\mathrm{Ln}(b')/\mathrm{max}(T_0,\Delta x) \numberthis \label{dx} \\
\beta_x &=\tan^{-1}\frac{d_x}{Q_k}, 0 < \beta_x < \pi/2 \\
  \Delta \xi_x &= \pi / \mathrm{max}(T_0,\Delta x) \numberthis \label{xi} \\
\xi_1 &= \left( \mathrm{Int} \left(P_k/\Delta \xi_x\right)+1 \right) \Delta \xi_x \numberthis \label{xi2} \end{align}
where Int$(\cdot)$ converts its argument to an integer via fractional truncation. Now parameterize the pre-extrapolation region integration path, for Re$[k_x]>0$, as
\begin{align} k_x &= \begin{cases} r\left(\cos\beta_x-i\sin\beta_x\right) &,  0<r<|Q_k+id_x| \\ r-id_x &, Q_k<r<P_k-Q_k \\ P_k-Q_k-id_x+r\left(\cos \beta_x+i\sin \beta_x \right) &, 0<r<|Q_k+id_x| \end{cases} \numberthis \label{kxReg}\\
 \frac{ \partial k_x}{\partial r} &= \begin{cases} \cos \beta_x-i\sin \beta_x  &,  0<r<|Q_k+id_x| \\ 1 &, Q_k<r<P_k-Q_k \\ \cos \beta_x+i\sin \beta_x &, 0<r<|Q_k+id_x| \end{cases} \end{align}
for the trapezoidal contour (used to integrate up to $k_x=P_k$) combined with a real-axis path to integrate within the section $P_k \leq k_x \leq \xi_1$. An analogous parameterization holds for the Re$[k_x]<0$ pre-extrapolation region path. Note that $\partial k_x/\partial r$ is independent of $k_x$ and thus can be computed prior to integration, unlike other commonly used detours\footnote{such as e.g. the half-sine-shaped detour \cite{mich1,sainath}}. This is the trapezoidal path's second benefit in addition to that mentioned in footnote \ref{ben1}.

Now we splice the regions $(0,P_k)$ and $(-P_k,0)$ each into $P$ regions, where $P$ is calculated as follows. First define
 \begin{equation}   d'=\mathrm{abs} \left( \frac{\partial \mathrm{e}^{ik_x\Delta x}}{\partial r}\bigg |_{r=|Q_k+id_x|} \right) =\sin(\beta_x)\Delta x\mathrm{e}^{d_x\Delta x}
\end{equation}
as the largest magnitude assumed by $\partial \mathrm{e}^{ik_x\Delta x}/ \partial r$ along the trapezoidal path, $c'$ as the user-defined maximum allowed magnitude change of $\mathrm{e}^{ik_x\Delta x}$ between two sampling points, and $T_1$ and $T_2$ as two user-defined parameters. Subsequently, define the quantities
 \begin{align} \Delta k_1 &= \mathrm{min}\left(\pi /( T_1 \mathrm{max}( \Delta x, \Delta z )),c'/d'\right) \\
 \Delta k_2 &= \pi /(T_1 \mathrm{max}( \Delta x, \Delta z )) \\
  N_{node,1} &=\mathrm{Int}\left(|Q_k+id_x| / \Delta k_1\right)+1  \\
    N_{node,2} &=\mathrm{Int}\left( (P_k-2Q_k)/\Delta k_2 \right)+1 \end{align}
which are used to yield $P_m$=Int(1+$N_{node,m}$/$T_2$) ($m$=1,2) with the corresponding final result $P=2P_1+P_2$. Note that this method of parameterizing the pre-extrapolation region path is empirical in nature and based on the pessimistic assumption of equidistant sampling \cite{sainath}.
\section{\label{ext}Extrapolation Region Path Revision}
The MWA, initially constructed in \cite{mosig2,mosig3} with further variants developed in \cite{mich2} and \cite{mosig1}, has also demonstrated the ability to accelerate convergence of infinite-range Fourier double-integrals in high-loss, planar-stratified environments containing anisotropic media \cite{sainath}. However, due to the highly oscillatory behavior of the mixed-domain integrand in integrals such as \eqref{SI1} arising from the Fourier kernels $\mathrm{e}^{ik_x \Delta x}$ and $\mathrm{e}^{ik_y \Delta y}$ when one has large $ \Delta x$ and $ \Delta y$ (resp.), the solution times (in our experience) became inordinately long. Therefore, it would be desirable to also deform the $k_x$ and $k_y$ plane ``extrapolation" region contours to lend additional \emph{exponential} decay via these two kernels, thereby dramatically accelerating convergence of the Fourier tail integrals and guaranteeing their \emph{absolute} convergence even in the ``worst-case" scenario $z-z'=0$. A cursory analysis reveals an apparent severe drawback, however: one can no longer employ the MWA, which was derived assuming a real axis integration path \cite{mosig1,mosig2,mosig3,mich2}. However, choosing a linear deformed path retains the MWA's algebraic convergence acceleration, as we show below\footnote{The MWA is retained for robustness in field solution acceleration; the mathematical and environmental constraints present typically prevent one from integrating along the ideal Constant Phase Path, as discussed below.}. For this analysis, take $k_x$ ($k_y$) as the inner (outer) integration variable.

We first exhibit the foundational relations needed to implement the complex-plane extension to a \emph{general} SLST followed by exhibiting the specific case arising from modeling the tail integral truncation error using the function family stipulated in the ``Mosig-Michalski Algorithm" (MMA) \cite{mich2,homeier}. Subsequently, we naively compute the optimal extrapolation region path detour angles without consideration for
\begin{enumerate}
\item the presence of critical points in the $k_x$ and $k_y$ planes and
\item \emph{two-dimensional} integrals, associated with wave propagation phenomena, imparting a transitory nature to these critical points in the $k_x$ plane (i.e. their locations now depend on the fixed $k_y$ value for which the $k_x$ integral is evaluated).
\end{enumerate}
To address the first concern, we pessimistically estimate the locations of critical points and reduce the $k_x$ plane departure angle of the deformed paths to ensure these features are not crossed. To address the latter concern, we (1) adjust the departure angles of the $k_y$ plane integration path and (2) partition the $k_x-k_y$ integration domain to ensure that the critical points
\begin{enumerate}
\item possess real parts with magnitude decaying as $|k_y|$ increases, leading to a bounded pre-extrapolation region, and
\item do not extend into the second/fourth quadrants, as this would require a) tracking their locations and b) adjusting the $k_x$ integration path, \emph{both} of which would become functions of $k_y$ and lead to a non-robust integration path.
\end{enumerate}
For simplicity, the analysis developing the complex-plane SLST generalization assumes isotropic planar layers\footnote{We justify this assumption based on previous analysis and results \cite{sainath}.}. Fix $k_y$ at some (generally complex) value $k_{y0}$ and assume $z-z' \geq 0$ \footnote{In Sections \ref{results}-\ref{conv} we demonstrate that the linear detour assures rapid convergence even in the regime $0 \leq |z-z'| \ll 1$.}; we see then that the inner integral of \eqref{SI1} writes as
\begin{equation}\label{int1} \int_{-\infty}^{\infty} \tilde{f}(k_x,k_{y0})\mathrm{e}^{ik_x\Delta x}\mathrm{e}^{ik_{y0}\Delta y}\mathrm{e}^{i\tilde{k}^+_z\Delta z} \mathrm{d}k_x \end{equation}
where $\tilde{k}_z^+$ is the up-going mode propagation constant, which for our time convention has positive imaginary part. Assuming $N$ extrapolation intervals are used \cite{mich2,sainath}, the linear path detour used in the $k_x$ integration path's extrapolation region is parameterized as
\begin{align} k_x &= \begin{cases} \xi_{1}+r_x\left(\cos{\gamma^+}+i\sin{\gamma^+}\right) &, 0 \leq r_x \leq N \Delta \xi_x^{+'}, \mathrm{Re}[k_x] > 0 \\ - \xi_{1}+r_x \left(\cos{\gamma^-}-i\sin{\gamma^-}\right) &, -N \Delta \xi_x^{-'} \leq r_x \leq 0, \mathrm{Re}[k_x] < 0 \end{cases} \numberthis \label{kxparam}\\
 \frac{\partial k_x}{\partial r_x} &= \begin{cases} \cos{\gamma^+}+i\sin{\gamma^+} &, 0 \leq r_x \leq N \Delta \xi_x^{+'}, \mathrm{Re}[k_x] > 0 \\ \cos{\gamma^-}-i\sin{\gamma^-} &, -N\Delta \xi_x^{-'} \leq r_x \leq 0, \mathrm{Re}[k_x] < 0 \end{cases} \numberthis \label{dkxparam} \end{align}
where one defines $\Delta \xi_x^{+'}=\Delta \xi_x / \cos{\gamma^+}$, $\Delta \xi_x^{-'}=\Delta \xi_x / \cos{\gamma^-}$, $\{ \gamma^+, \ \gamma^- \} \geq 0$, and $r_x$ as real-valued. $\xi_{1}$ is assumed large enough to ensure that we have sufficiently detoured past any critical points near the real axis \cite{mosig1}. Now recall that plane wave propagation in a homogeneous, unbounded, isotropic medium with wave number $k$ is governed by the dispersion relation $(\tilde{k}^+_z)^2=(\tilde{k}^-_z)^2=(k'+ik'')^2-k_x^2-k_{y0}^2$  \cite{chewch1}. For large $|k_x|$ this relation becomes\footnote{In arriving at \eqref{dispasym}, the large-$|k_x|$ form of the dispersion relation, the proper square root sign is taken to assure exponential decay of the Fourier kernel $\mathrm{e}^{i\tilde{k}^+_z\Delta z}$ in accordance with the radiation condition.}
\begin{equation} \label{dispasym}
\tilde{k}^+_z \to i \begin{cases} \xi_1+r_x \left( \cos{\gamma^+}+i\sin{\gamma^+} \right) &, \mathrm{Re}[k_x] > 0 \\
\xi_1- r_x\left( \cos{\gamma^-}-i\sin{\gamma^-} \right) &, \mathrm{Re}[k_x]<0 \end{cases}
 \end{equation}
Next, assuming that (asymptotically) $\tilde{f}(k_x,k_{y0}) \to k_x^q\sum_{m=0}^\infty \tilde{a}_m(k_{y0})\tilde{\phi}_m(k_x)$ \cite{mich2,homeier}, one can substitute this series expression into the extrapolation region section of \eqref{int1} to obtain
\begin{equation}\label{int1b} \int_{\mathrm{ext}} k_x^q \sum_{m=0}^\infty \tilde{a}_m\tilde{\phi}_m(k_x) \mathrm{e}^{ik_x\Delta x}\mathrm{e}^{ik_{y0} \Delta y}\mathrm{e}^{i\tilde{k}^+_z\Delta z} \mathrm{d}k_x \end{equation}
where ``ext" denotes the $k_x$ plane extrapolation region integration path section and the $\{\tilde{\phi}_m \}$ comprise a family of functions used to asymptotically model $\tilde{f}(k_x,k_{y0})$ and the truncation error (discussed below) \cite{homeier}. Setting $t_o^+=\cos{\gamma^+}+i\sin{\gamma^+}$, $t_o^-=\cos{\gamma^-}-i\sin{\gamma^-}$, $l^+ = t_o^+ \mathrm{e}^{ik_{y0} \Delta y}\mathrm{e}^{\xi_1(i\Delta x-\Delta z)}$, and $ l^- = t_o^- \mathrm{e}^{ik_{y0} \Delta y}\mathrm{e}^{-\xi_1(i\Delta x+\Delta z)}$, \eqref{int1b} becomes the union of \eqref{int1ci} and \eqref{int1cii}:
  \begin{align} I_{\mathrm{ext}}^+&= l^+\sum_{m=0}^\infty \int_{0}^\infty  \tilde{a}_m\tilde{\phi}_m(\xi_{1}+r_xt^+_o) (\xi_{1}+r_xt^+_o)^q\mathrm{e}^{r_xt^+_o(i\Delta x-\Delta z)}\mathrm{d}r_x \numberthis \label{int1ci} \\
  I_{\mathrm{ext}}^-&= l^-\sum_{m=0}^\infty \int_{-\infty}^0  \tilde{a}_m\tilde{\phi}_m(-\xi_{1}+r_x t^-_o)(-\xi_{1}+r_xt^-_o)^q\mathrm{e}^{r_x t^-_o(i\Delta x+\Delta z)} \mathrm{d}r_x \numberthis \label{int1cii}  \end{align}
with the respective truncation error integrals of \ref{int1ci}-\ref{int1cii} manifesting as
\begin{align} I_{\mathrm{tr}}^+ &= l^+\sum_{m=0}^\infty \int_{ N\Delta \xi_x^{+'}}^\infty  \tilde{a}_m\tilde{\phi}_m(\xi_{1}+r_xt^+_o)(\xi_{1}+r_xt^+_o)^q\mathrm{e}^{r_xt^+_o(i\Delta x-\Delta z)}\mathrm{d}r_x \numberthis \label{int1di}\\
 I_{\mathrm{tr}}^- &=l^-\sum_{m=0}^\infty \int_{-\infty}^{-N\Delta \xi_x^{-'}}  \tilde{a}_m\tilde{\phi}_m(-\xi_{1}+r_x t^-_o)(-\xi_{1}+r_xt^-_o)^q\mathrm{e}^{r_x t^-_o(i\Delta x+\Delta z)} \mathrm{d}r_x \numberthis \label{int1dii} \end{align}
Performing a change of variables on \eqref{int1di}-\eqref{int1dii} subsequently yields the following relations
\begin{align} I_{\mathrm{tr}}^+ &=l^+ \sum_{m=0}^\infty \int_0^\infty  \tilde{a}_m\tilde{\phi}_m(\xi_{1}+(s+N\Delta \xi_x^{+'})t^+_o)(\xi_{1}+(s+N\Delta \xi_x^{+'})t^+_o)^q\mathrm{e}^{(s+N\Delta \xi_x^{+'})t^+_o(i\Delta x-\Delta z)}\mathrm{d}s \numberthis \label{int1ei} \\
I_{\mathrm{tr}}^- &=l^- \sum_{m=0}^\infty \int_0^\infty  \tilde{a}_m\tilde{\phi}_m(-\xi_{1}-(s+N\Delta \xi_x^{-'})t^-_o)(-\xi_{1}-(s+N\Delta \xi_x^{-'})t^-_o)^q\mathrm{e}^{-(s+N\Delta \xi_x^{-'})t^-_o(i\Delta x+\Delta z)}\mathrm{d}s \numberthis \label{int1eii} \end{align}
Next, one evaluates \eqref{int1ei}-\eqref{int1eii} for $M+1$ different values of $N$ (e.g. $N=1,2,...,M+1$), truncates these $M+1$ relations after the $m=(M-1)$ error series term (i.e. retain the first $M$ series terms), defines the $m$th truncation error series coefficient pair as $\{ \tilde{c}^+_m,\tilde{c}^-_m \}$ (e.g. see \eqref{int1gia}-\eqref{int1giia} below), and solves the corresponding $(M+1)$-order system to estimate $I^+_{\mathrm{ext}}$ and $I^-_{\mathrm{ext}}$. This procedure represents the complex-plane SLST generalization, applicable to the sequence of $M+1$ successive ``cumulative tail integral" estimates \cite{sainath}, to accelerate evaluation of $I^+_{\mathrm{ext}}$ and $I^-_{\mathrm{ext}}$.

Let us now examine the specific case of modeling $\tilde{f}(k_x,k_{y0})$ using the family of $(N,m)$-parameterized functions $\tilde{\phi}_m=\tilde{\phi}^+_m(N,s)=(\xi_{1}+(s+N\Delta \xi_x^{+'})t^+_o)^{-m}$ for Re$[k_x]>0$ and $\tilde{\phi}_m=\tilde{\phi}^-_m(N,s)=(-\xi_{1}-(s+N\Delta \xi_x^{-'})t^-_o)^{-m}$ for Re$[k_x]<0$. Performing Maclaurin expansions of the $\{\tilde{\phi}^+_{m-q} \}$ and $\{ \tilde{\phi}^-_{m-q} \}$, retaining only their respective zeroth-order expansion terms, setting $\omega^+_N=\mathrm{exp}(-N\Delta \xi_x^{+'}\left(\cos{\gamma^+}\Delta x+\sin{\gamma^+}\Delta z\right))$ and $\omega^-_N=\mathrm{exp}(-N\Delta \xi_x^{-'}\left(\cos{\gamma^-}\Delta x+\sin{\gamma^-}\Delta z\right))$, and defining
\begin{align} \omega^{+'}_N &=\omega^+_N\tilde{\phi}^+_{-q}(N,0), \ \tilde{c}_m^+ \sim l^+\int_0^\infty  \tilde{a}_m\mathrm{e}^{st^+_o(i\Delta x-\Delta z)}\mathrm{e}^{iN\Delta \xi_x^{+'}( \cos{\gamma^+} \Delta x- \sin{\gamma^+} \Delta z )}\mathrm{d}s \numberthis \label{int1gia} \\
\omega^{-'}_N &=\omega^-_N\tilde{\phi}^-_{-q}(N,0), \ \tilde{c}_m^- \sim l^- \int_0^\infty  \tilde{a}_m\mathrm{e}^{-st^-_o(i\Delta x+\Delta z)}\mathrm{e}^{-iN\Delta \xi_x^{-'}(\cos{\gamma^-} \Delta x - \sin{\gamma^-} \Delta z )}\mathrm{d}s \numberthis \label{int1giia} \end{align}
yields a pair of expressions capturing the dominant behavior of the truncation error $I_{\mathrm{tr}}^+ \cup I_{\mathrm{tr}}^-$:
\begin{align} I_{\mathrm{tr}}^+ &\sim \omega^{+'}_N\sum_{m=0}^\infty \tilde{c}_m^+\tilde{\phi}^+_{m}(N,0)  \numberthis \label{int1gi} \\
I_{\mathrm{tr}}^- &\sim \omega^{-'}_N \sum_{m=0}^\infty \tilde{c}_m^-\tilde{\phi}^-_{m}(N,0) \numberthis \label{int1gii} \end{align}
which comprises the complex-path generalization, as concerning infinite-range Fourier integrals, to the error expression developed in Section 2 of \cite{mich2}. The corresponding truncation error expressions associated with F-H transforms like \eqref{SI2} follow in analogous fashion.

With the foundational expressions available, we now seek to maximize exponential convergence acceleration of \eqref{int1b} through a suitable choice of the detour departure angles $\gamma^+$ and $\gamma^-$. Differentiating the exponent expressions $\left( \cos{\gamma^+}\Delta x+\sin{\gamma^+}\Delta z \right)$ and $\left( \cos{\gamma^-}\Delta x+\sin{\gamma^-}\Delta z \right)$ with respect to $\gamma^+$ and $\gamma^-$ (resp.) and setting the resulting expressions equal to zero leads us to initially (naively) choose $\gamma^+=\gamma^-=\mathrm{tan}^{-1}\left(\Delta x / \Delta z\right)$\footnote{For F-H integrals, use the asymptotic form of the Hankel function and replace $\Delta x$ with $|\rho-\rho'|$ when computing $\gamma^+=\gamma^-$.}, which (asymptotically) corresponds to the path of most rapid exponential decay\footnote{For reasons discussed in \cite{davies}, this path appears similar to, but is \emph{not} always necessarily, the SDP.} or (equivalently) the Constant Phase Path (CPP). This detour angle choice can be likened to a compromise between the so-called ``$z$-transmission representation" and ``radial transmission representation" \cite{mosig6} of the space-domain field, which were discussed therein in the context of F-H and F-B transforms. Identical expressions hold for the $k_y$ plane detour departure angles $\alpha^+$ and $\alpha^-$. Next we consider the detour constraints imposed by these two phenomena.

To this end, first define the branch point as $k_a^2=k^2-k_{y0}^2$ and temporarily assume that the $k_y$ path was chosen so that $k_x$ plane critical points neither manifest in the second/fourth quadrants nor migrate towards Re[$k_x$]$=\pm \infty$ with increasing $|k_y|$. Recalling the effective refractive indices $\{ \sqrt{\epsilon_{pm}\mu_{pn}} \ \}$ ($m,n=1,2,3$) for layer $p$\footnote{These are used to compute, but are \emph{not} the same as, the global effective refractive index $n^+$ mentioned above (see \cite{sainath}).} and how we subsequently computed $\xi_1$ \cite{sainath}, set $\gamma_{m,n,p}$ equal to either (1) the angle between $\xi_1$ and the $p$th layer's $(m,n)$th ``effective wave number"\footnote{After coercing the wave number's real part to be positive, if need be.} $k_{m,n,p}=k_o \sqrt{\epsilon_{pm}\mu_{pn}}$, if $\mathrm{Re}[k_{m,n,p}] \geq \xi_1$, or (2) $\pi/2$ if $0 \leq \mathrm{Re}[k_{m,n,p}] < \xi_1$. Then, $\gamma^+$ is updated as $\gamma^+=\mathrm{min}(\gamma^+, \ \{ \gamma_{m,n,p} \})$. No critical points are located in the second/fourth quadrants by assuming (for simplicity) the absence of ``double-negative"/meta-material and active/gain media. Therefore, we do not have to constrain $\gamma^-$. However, these calculations can be readily adjusted to appropriately constrain both $\gamma^+$ and $\gamma^-$ if such media are present so that our assuming their absence represents a trivial constraint in our methodology.

Now we justify the assumptions above about the $k_y$ path, and constrain it to avoid the two issues stated earlier regarding two-dimensional integral transforms arising as the solution to wave-dynamics problems in planar-stratified environments lacking azimuthal symmetry. To this end, for some arbitrary $k_y$ value along the $k_y$ plane integration path first expand the branch point $k_a^2$ as\footnote{One can verify that the $k_x$ plane discrete poles will exhibit similar behavior as the branch points \cite{chewch2}. Therefore, our analysis based on examining the branch point's behavior sheds analogous insight into the behavior of the poles.}
\begin{equation}\label{kasq} k_a^2 = \left[(k^{'2}-k^{''2})-(k_y^{'2}-k_y^{''2})\right]+2i\left[k'k''-k_y'k_y''\right]\end{equation}
 and recall that the radiation branch cut is jointly defined by the conditions Im$[\tilde{k}_z^2]=0$ and Re$[\tilde{k}_z^2]>0$ \cite{chewch2}. To ensure that critical points in the first (third) quadrant of the $k_x$ plane do not migrate towards Re[$k_x$]=$+\infty$ (Re[$k_x$]=$-\infty$) for large $|r_y|$\footnote{$r_y$ is the $k_y$ plane dual of $r_x$, exhibited earlier.}, one must ensure that asymptotically Re$[k_a^2] \to -\infty$ as $|r_y| \to \infty$. Observing the real part of \eqref{kasq}, we see that one must constrain $\alpha^-$ and $\alpha^+$ to the interval $0 \leq \{ \alpha^-,\alpha^+ \} \leq \pi/4$. Furthermore, to prevent critical points from migrating into the second/fourth $k_x$ plane quadrants, we require that Im$[k_a^2] \geq 0$ as $|r_y| \to \infty$. Observing the imaginary part of \eqref{kasq} and noting in the region $\{ k_y' > 0 \ \cup \ k_y''>0 \}$ that $k_y'k_y''>0$, a cursory analysis suggests that one cannot safely choose a non-zero value of $\alpha^+$ without risking this migration, which would force one to dynamically re-define the $k_x$ integral's pre-extrapolation region path, now a function of $k_y$, to ensure that one (1) encloses all the quadrant one critical points that migrated into quadrant four while (2) avoiding the encirclement of quadrant three critical points that migrated into quadrant two.

\begin{figure}[H]
    \centering
    \vspace{-10pt}
    \includegraphics[width=3.5in]{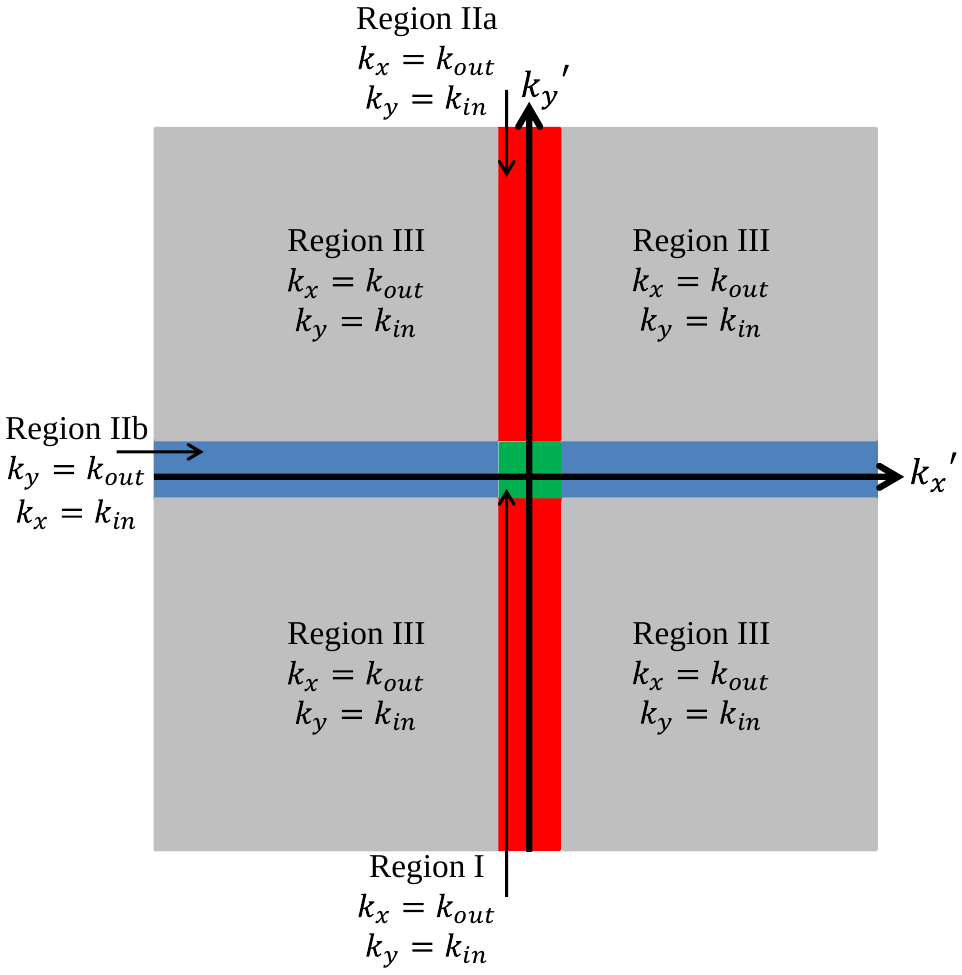}
    \vspace{-5pt}
    \caption{\small Depiction of the proposed integration domain partition scheme to ensure absolute convergence of Fourier double-integrals such as \eqref{SI1} and \eqref{SI3}. $k_{out}$ and $k_{in}$ represent the outer and inner integration variables (resp.).}
    \label{IntScheme}
    \end{figure}

 As a result, it appears that one must set the additional, more restrictive constraint $\alpha^+=\mathrm{min}(0,\pi/4)=0$, which in theory may lead to an outer integral exhibiting monotonic-divergent behavior when $0 \leq |z-z'| \ll 1$ \cite{mosig1}. However, this limitation can be overcome via clever partition of the two-dimensional integration domain; see Figure \ref{IntScheme}, which summarizes the proposed partition. Integrating first over Regions I, IIa, and IIb in Figure \ref{IntScheme} followed by integrating in Region III, which encompasses the intersection of the $k_x$ and $k_y$ plane extrapolation regions, renders the Region III integration's result \emph{immune} to the migration of critical points into the second/fourth quadrants. This is because one had already stipulated a domain partitioning and completed integration over Regions I, IIa, and IIb.

We conclude that so long as one conforms to the restrictions $0 \leq \{ \alpha^-,\alpha^+ \} \leq \pi/4$ and $\gamma^+=\mathrm{min}(\mathrm{tan}^{-1}\left(\Delta x / \Delta z\right), \\ \{ \gamma_{m,n,p} \})$, one can detour in \emph{all four} spectral ``quadrants" $\{ k_x'>0 \cup k_y'>0 \}$, $\{ k_x'<0 \cup k_y'>0 \}$, $\{ k_x'>0 \cup k_y'<0 \}$, and $\{ k_x'<0 \cup k_y'<0 \}$ (see Figure \ref{IntScheme}) through which the stipulated integration path proceeds\footnote{Due to our assuming $x-x'=y-y' \geq 0$, it is implicitly understood that in all four ``quadrants" Im[$k_x] \geq 0$ and Im[$k_y] \geq 0$ (excepting the minor pre-extrapolation region detour made into quadrant four).}. Indeed, our proposed partition of the $k_x-k_y$ integration domain ensures that for \emph{any} $\bold{r}-\bold{r}' \neq \bold{0}$ geometry, the double-integral \eqref{SI1} exhibits \emph{absolute-convergent} behavior in the classical/Riemann sense and thus, by Fubini's theorem \cite{fubini}, can be evaluated using an iterated integral whose solution is independent of the order of integration.

This discussion also brings to light the benefit of our starting assumption in this analysis, made at the end of Section \ref{intro}, that $|x-x'|=|y-y'| \geq 0$: a compromise is reached that ensures exponential-cum-algebraic convergence of both the $k_x$ \emph{and} $k_y$ integrals throughout the integration domain. As an alternative we could have, for example, performed an azimuthal rotation such that $|x-x'|=0$ and $|y-y'|=|\rho-\rho'|$ to maximize convergence acceleration of the $k_y$ integral. However, when $0 \leq |z-z'| \ll 1$, the $k_x$ integral may exhibit \emph{monotonic-divergent} behavior. In contrast to oscillatory-divergent behavior \cite{mosig1}, the MWA variants (including the generalized version developed herein) cannot curb monotonic-divergent behavior due to the lack of oscillations that must be present for the MWA to ``average out" the oscillatory-divergent sequence of cumulative tail integral estimates to obtain a final, convergent result.

Note that for integration in Regions IIa, IIb, and III in Figure \ref{IntScheme}, one performs a separate integration and extrapolation of the individual $k_x$ and/or $k_y$ half-tail integral sections. This is in contrast to the method developed in \cite{sainath} wherein we folded the half-tail integrals in the $\pm \mathrm{Re}[k_x]$ half-planes to yield a cosine or sine oscillatory kernel, based on assuming spectral symmetry in the environment's plane wave reflection/transmission properties, prior to performing tail integral extrapolation along the positive Re[$k_x$] axis (and similarly for the $k_y$ plane). Our present method, in bending both half-tail $k_x$ paths into the upper-half $k_x$ plane, forbids such folding due to the now-absent lack of reflection symmetry (about the Im[$k_x$] axis) with respect to the two halves of the extrapolation region path. The resulting penalty paid in using the complex-plane MWA manifests in having to use twice the number of weight sets versus when one can perform half-tail integral folding followed by cumulative tail integral sequence extrapolation, leading to increased memory requirement and computation time in regards to procuring the MWA weight sets. As a practical consideration, then, we wish to reduce the number of extrapolation weight sets that must be evaluated\footnote{Nominally, there are twelve weight sets one must pre-compute and store to implement the complex-plane MWA: three field components, each with differing combinations of $(k_x,k_y)$ monomial power dependencies, multiplied by up to four distinct extrapolation region detour angles $\alpha^+, \alpha^-, \gamma^+$, and $\gamma^-$.}. To this end, we take two steps to halve this number to the six weight sets originally required when performing folding followed by extrapolation. First, we set $\alpha^+=\gamma^+$ and $\alpha^-=\gamma^-$. Second, we make the approximation (for each field component) that the asymptotic monomial power dependence on both $|k_x|$ and $|k_y|$ \cite{chewch1,chewch7} equals the average monomial power dependencies on $k_x$ and $k_y$. For example, if we determine the integrand for one field component has asymptotic monomial dependencies of $\mathcal{O}(k_x^{q_1})$ and $\mathcal{O}(k_y^{q_2})$, then we take (as both our $k_x$ and $k_y$ monomial dependence factors) $q_o=\mathrm{Nint}((q_1+q_2)/2)$, where Nint($\cdot$) converts its argument to an integer via rounding. Furthermore, to ensure stability of the accelerator weight expressions and minimize aliasing effects due to inordinately long extrapolation region intervals, we neglect the integrand oscillation due to $\Delta z$ in the exponential kernels of \eqref{int1gia}-\eqref{int1giia}\footnote{Stability and aliasing considerations also motivated our choice of the break-point spacings $\Delta \xi_x$, $\Delta \xi_x^{+'}$, and $\Delta \xi_x^{-'}$.}. This allows one to update the truncation error estimates $\{ \omega_N^{+'},\omega_N^{-'} \}$ as $\omega_N^{\pm'}=(-1)^N\omega_N^{\pm'}$ when $\Delta x>T_0$ \cite{mich2}.

Beyond the concern of weight computation stability, we also ignore the phase variation associated with $\Delta z$ due to $\Delta z$, in general, being ill-defined. Indeed, in (1) an anisotropic homogeneous environment or (2) a stratified environment containing isotropic and/or anisotropic media, several phenomena typically obfuscate a univocal, clear definition for the effective longitudinal distance traversed by the characteristic plane wave fields when traveling from $\bold{r}'$ in layer $M$ to $\bold{r}$ in layer $L$. These are (1) multi-bounce within slab layers, (2) the layer and (for anisotropic media) mode dependence of the longitudinal propagation constants, (3) interface reflections in layer $L$ causing both up-going and down-going modal fields (four total modes in general) to contribute to the observed field at $\bold{r}$, and (4) inter-mode coupling at the interfaces. In fact these four considerations, along with the inherently asymptotic nature of the CPP parameterization and the constraints associated with critical points/two-dimensional integrals addressed above, lead one in practice to not integrate exactly along the CPP. As a result, one typically finds the integrands of extrapolation-region integrals still exhibiting undesirable residual oscillation due to the complex exponential factors. While, for $\Delta x \geq 0$, one still \emph{always} has a non-zero detour angle for both the $k_x$ and $k_y$ extrapolation region paths\footnote{Recall from \cite{sainath} that the pre-extrapolation region serves to detour around those critical points within a certain distance from the real axis. Therefore, the presence of critical points cannot force $\gamma^+$ and $\gamma^-$ to equal zero exactly.}, these practical considerations are what demand the inclusion of an algebraic convergence accelerator like the MWA that exactly acts upon the very types of oscillatory integrals that will typically result. Therefore, while one does not typically realize the ideal situation of maximized exponential convergence acceleration (the strongest acceleration theoretically available here outside of the SDP), we largely mitigate this pitfall with the robust algebraic acceleration afforded by the MWA, which is agnostic to the environment/source-observer scenario (so long as $\Delta x >0$). Indeed, for an order-$N$ MWA method used (see below) one realizes a reduction in truncation error between $\mathcal{O}(k_x^{-N}$) and $\mathcal{O}(k_x^{-2N}$) \cite{mich2}.
\section{\label{weightcomp}Revised Accelerator Weight Computation}
The MWA, both in its form as the MMA \cite{mich2,homeier} and its more recent variant the new/``revisited" MWA \cite{mosig1}, each offer different, desirable attributes. The latter version offers a straightforward methodology to unambiguously define arbitrary-order accelerator weight sets and recursively compute higher-order weight sets upon demand. While we showed previously \cite{sainath} for the MMA how one can reduce the FLOP\footnote{Floating Point Operations} count involving the cumulative integrals themselves, the weight computations (1) depended on whether the series of successive extrapolation region sub-interval integrals exhibited oscillating or monotone behavior \cite{mich2,sainath}, and (2) the FLOP count to compute the weight sets rapidly grows for successive weight sets, placing a practical limit on obtainable accuracy in the weights (and thus the estimated tail integral) due to roundoff error accumulation in the computed weights. On the other hand, the computation of the new MWA weight sets is (1) a numerically unstable process rapidly leading to numerical overflow (when using the form exhibited in \cite{mosig1}) and (2) directly linked to procuring the estimated tail integral \cite{mosig1}, which is the solution to a highly ill-conditioned linear system (shown in Section \ref{formulations}), which previously led us to use the MMA in \cite{sainath}. Nevertheless, both flavors of MWA offer useful mathematical developments for the weights that are couched in the framework of SLST, using a family of functions in a series representation to model the spectral portion of the mixed-domain Green's Function\footnote{i.e. the fundamental spectral kernel embedded in the integrands of \eqref{SI1}-\eqref{SI4}.} and resulting tail integral truncation error. For the MMA, the proposed series \cite{mich2}
\begin{equation}\label{GFapp} \tilde{f}(k_x,k_{y0}) \sim  k_x^q \sum_{m=0}^{\infty} \frac{\tilde{a}_m(k_{y0})}{k_x^m} \end{equation}
is intuitive in its form, and we confirm below the validity (and in fact \emph{optimality}) of using this approximating series by a straightforward mathematical analysis\footnote{When assuming the spectral portion of the integrands in \eqref{SI1}-\eqref{SI4} asymptotically behave as monomial powers of $k_x$ and $k_y$, the new MWA's error-modeling functions reduce to this function family too \cite{mosig1}.}. It will be shown that this optimality arises due to the error modeling series (see \eqref{int1gi}-\eqref{int1gii} above and \cite{mich2}) being entirely consistent with the closed-form expression of the truncation error both in the absence and (via linear superposition) presence of stratified inhomogeneity. The practical consequence of this function family's modeling optimality manifests in minimizing the number of cumulative tail integral estimates required to accurately estimate $I^+_{\mathrm{ext}}$ and $I^-_{\mathrm{ext}}$, as demonstrated in Section \ref{conv}.

In summary, we seek a revised, complex-plane MWA that combines the best aspects of both the MMA \cite{mich2} and more recent MWA variant \cite{mosig1}, in tandem with incorporating the added exponential convergence acceleration afforded by bending the extrapolation region integration path, to effect robust and powerful field solution convergence acceleration. To this end, in this section we (1) analyze and justify using \eqref{GFapp} as the approximating series for $\tilde{f}(k_x,k_{y0})$ and (2) exhibit and compare two proposed formulations for implementing the complex-plane generalization of the MWA, using the new ``remainder estimates" $\{\omega_N^{+'},\omega_N^{-'}\}$ \cite{mich2} and the asymptotic series expansion \eqref{GFapp}.
\subsection{The Optimal Error-Modeling Function Family}
Herein we examine the inner spectral ($k_x$) integral for some fixed $k_y=k_{y0}$ in the region Re$[k_x]>0$. Furthermore, assume $\tilde{f}(k_x,k_{y0})$ has an asymptotic $k_x$ monomial dependence of $k_x^q$ \cite{mosig1,mich2}. One then has the asymptotic truncation error
\begin{equation}\label{int2} I_{\mathrm{tr}}^+=\omega^+_N I_{\mathrm{tr}}^{+'}=\omega^+_N \int_0^\infty  \left[\xi_1+(s+N\Delta \xi_x^{+'})t_o^+\right]^q \mathrm{e}^{st^+_o(i\Delta x-\Delta z)}\mathrm{d}s \end{equation}
which just equals \eqref{int1ei} with the asymptotic series expansion for $\tilde{f}(k_x,k_{y0})$ replaced by the dominant series term $k_x^q$. Next note that $I_{\mathrm{tr}}^{+'}$ has a closed-form, convergent solution for $\mathrm{Re}[t^+_o(i\Delta x-\Delta z)]<0$; setting $\tilde{a}=\tilde{a}(N)=\tilde{\phi}^+_0(N,0)=\xi_1+t_o^+N\Delta \xi_x^{+'}$ and $\chi=-t^+_o(i\Delta x-\Delta z)$, one obtains
\begin{equation}\label{tail1} I_{\mathrm{tr}}^{+}=\omega_N^+I_{\mathrm{tr}}^{+'}= \omega_N^+\begin{cases} 1 / \chi &, q=0 \\ (t_o^++\tilde{a}\chi)/\chi^2 &, q=1 \\ (2t_o^{+2}+2\tilde{a}t_o^+\chi+\tilde{a}^2\chi^2)/\chi^3 &, q=2 \end{cases} \end{equation}
and so on for other values of $q$ \footnote{\label{super}For the multi-layered scenario, use \eqref{int1ei} and invoke superposition of the closed-form results for different monomial powers $q-m$ ($m=0,1,2,...$).}. Examining the asymptotic limit for these three illustrative cases, we find:
\begin{equation}\label{chiasymp} \lim_{|\tilde{a}| \to \infty} I_{\mathrm{tr}}^{+'}= \begin{cases} 1/\chi &, q=0 \\ \tilde{a}/\chi &, q=1 \\ \tilde{a}^2/\chi &, q=2 \\ \tilde{a}^q/\chi &, q \in \mathcal{N}^+\end{cases} \end{equation}
where $\mathcal{N}^+$ represents the set of positive natural numbers. Similarly, one expects that the reflected/transmitted field terms will also have an asymptotic monomial dependence \cite{mosig1}. For example, consider a two-layer, planar-stratified environment containing isotropic media. The TE$_z$/TM$_z$ reflection and transmission coefficients for a plane wave, incident from half-space number one upon half-space number two, write as \cite{chewch2}
\begin{equation}\label{TR2} R_{\mathrm{TM}}=\frac{\epsilon_{2,r}\tilde{k}^+_{1,z}-\epsilon_{1,r}\tilde{k}^+_{2,z}}{\epsilon_{2,r}\tilde{k}^+_{1,z}+\epsilon_{1,r}\tilde{k}^+_{2,z}}, \ R_{\mathrm{TE}}=\frac{\mu_{2,r}\tilde{k}^+_{1,z}-\mu_{1,r}\tilde{k}^+_{2,z}}{\mu_{2,r}\tilde{k}^+_{1,z}+\mu_{1,r}\tilde{k}^+_{2,z}}, \
T_{\mathrm{TM}}=1+R_{\mathrm{TM}}, \
T_{\mathrm{TE}}=1+R_{\mathrm{TE}}\end{equation}
where
\begin{equation}\label{TR1} (\tilde{k}^+_{m,z})^2 =(\tilde{k}^-_{m,z})^2= k_o^2\epsilon_{m,r}\mu_{m,r}-k_{\rho}^2 \ \ (m=1,2), \ \lim_{|k_{\rho}| \to \infty} \tilde{k}^+_z = ik_{\rho} \end{equation}
Indeed, we see that for large $|k_{\rho}|$ the reflection/transmission coefficients have a monomial power dependence $\mathcal{O}(k_{\rho}^0)$.

Pulling out the dominant monomial term $\tilde{a}^q$ in \eqref{tail1}, setting the $N$th cumulative remainder estimate as $\omega_N^{+'}=\omega_N^{+}\tilde{a}^q$, and invoking superposition (see footnote \ref{super}), one now has $\omega_N^{+}I_{\mathrm{tr}}^{+'}=\omega_N^{+'}\sum_{m=0}^{q}\tilde{b}^+_m/\tilde{a}^m$, which recovers the dominant contribution to the complex-path extension \eqref{int1gi} of the error expression derived (using \eqref{GFapp}) in \cite{mich2}\footnote{The coefficients $\{\tilde{b}^+_m\}$, however, are computed \emph{exactly}; contrast this to the $\{ \tilde{c}_m^+ \}$ of \eqref{int1gi}.}. The same procedure shown above, using instead $t_o^-$, $\tilde{a}=\tilde{\phi}^-_0(N,0)=-\xi_1-t_o^-N\Delta \xi_x^{-'}$, and $\chi=t^-_o(i\Delta x+\Delta z)$, can be repeated for the Re$[k_x]<0$ tail integral to obtain a dual set of expressions that recover \eqref{int1gii}. Based on this analysis, when proposing two revised MWA methods we will use the $(N,m)$-parameterized function families $\{\tilde{\phi}^+_m(N)= \tilde{\phi}^+_m(N,0) \}$ and $\{ \tilde{\phi}^-_m(N)= \tilde{\phi}^-_m(N,0) \}$ to model the tail integral truncation error.
\subsection{\label{formulations}Two Proposed Formulations}
For the first formulation we take inspiration from \cite{mosig1}. To this end, for the Re$[k_x]>0$ tail integral first define $I_{N}^{+'}$ and $I_{N+1}^{+'}$ as two input cumulative tail integral estimates and the under-determined linear system, with respect to which the non-truncated tail integral $I^+_{\mathrm{ext}}$ is defined, as \cite{mosig1,homeier}
 \begin{align} I^+_{\mathrm{ext}} &=I_N^{+'}+ \omega_N^{+'}\sum_{m=0}^{\infty} \frac{\tilde{b}^+_m}{\tilde{a}(N)^m} \numberthis \label{linsys}\\  I^+_{\mathrm{ext}} &=I_{N+1}^{+'}+ \omega_{N+1}^{+'}\sum_{m=0}^{\infty} \frac{\tilde{b}^+_m}{\tilde{a}(N+1)^m}  \numberthis \label{linsys2}\end{align}
 whose equations are subsequently truncated after the $m=0$ term \cite{mosig1}. This truncation yields a second-order linear system solved for an improved estimate $I^{+(2)}_N$ of $I^+_{\mathrm{ext}}$ that is free of the $\tilde{a}^0$ term in its truncation error series \cite{mosig1,mich2,homeier}:
 \begin{equation} \eta_N^{+(1)}=-\frac{\omega_{N}^{+'}}{\omega_{N+1}^{+'}}, \ \ I^{+(2)}_N=\frac{I_N^{+'}+\eta_N^{+(1)}I_{N+1}^{+'}}{1+\eta_N^{+(1)}}=
I^+_{\mathrm{ext}}+\omega_N^{+'}\frac{\sum_{m=1}^{\infty}\tilde{b}^+_m \left[\tilde{a}(N+1)^{-m}-\tilde{a}(N)^{-m}\right]}{1+\eta_N^{+(1)}}\end{equation}
 Similarly, using $M+1$ ($M=1,2,...$) cumulative tail integral estimates $\{I_{1}^{+'}, I_{2}^{+'}, ..., I_{M+1}^{+'} \}$ to eliminate the first $M$ terms of $\sum_{m=0}^{\infty}\tilde{b}^+_m/\tilde{a}^m$, one has for the $P$th truncated linear equation ($P=1,2,...,M+1$) $I^{+(M+1)}_1 =I_{P}^{+'}+\omega_{P}^{+'}\sum_{m=0}^{M-1} \tilde{b}^+_m/\tilde{a}(P)^m$. Subsequently, one procures the weights via solving the associated order-($M+1$) linear system for the best $I^+_{\mathrm{ext}}$ estimate (i.e. $I^{+(M+1)}_N$), whose solution implicitly contains the expressions for the weights \cite{mosig1}. However, obtaining all desired weight tier sets by directly solving the associated linear systems (1) is very costly and (2) possibly exacerbates weight accuracy degradation due to the poor conditioning of these systems (see below). Instead, one can obtain closed-form solutions to the weight sets using the methodology outlined in \cite{mosig1} as adapted to our choice of (1) error-modeling functions $\{ \tilde{a}(N)^{-m} \}$ and (2) truncation error estimates $\{ \omega_N^{+'} \}$. In \cite{mosig1} it was assumed that $\tilde{f}(k_x,k_{y0})$ asymptotically exhibited a monomial power dependence of the form $Ck_x^q$ ($C$ being some constant), with the obvious consequence that $\partial^n \tilde{f}(k_x,k_{y0})/\partial^n k_x$ corresponds to a new function asymptotically behaving as $\sim k_x^{q-n}$ \footnote{Examining \cite{chewch2,chewch7} confirms the asymptotic monomial power dependence of $\tilde{f}(k_x,k_{y0})$.}. Rearranging the order-($M+1$) linear system thus yields a similar (but not yet identical) system to equation (22) in \cite{mosig1}:
\begin{equation}\label{linsys3}
\begin{bmatrix} -1/\omega_1^{+'} & 1 & \tilde{a}(1)^{-1} & \cdots & \tilde{a}(1)^{-(M-1)} \\ -1/\omega_2^{+'} & 1 & \tilde{a}(2)^{-1} & \cdots & \tilde{a}(2)^{-(M-1)} \\
\vdots & \vdots & \vdots & \vdots & \vdots \\
-1/\omega_{M+1}^{+'} & 1 & \tilde{a}(M+1)^{-1} & \cdots & \tilde{a}(M+1)^{-(M-1)} \end{bmatrix}\begin{bmatrix} I^{+(M+1)}_1 \\ \tilde{b}^+_o \\ \tilde{b}^+_1 \\ \vdots \\ \tilde{b}^+_{M-1}\end{bmatrix}=-\begin{bmatrix} I_{1}^{+'}/\omega_1^{+'} \\ I_{2}^{+'}/\omega_2^{+'} \\ I_{3}^{+'}/\omega_3^{+'} \\ \vdots \\ I_{M+1}^{+'}/\omega_{M+1}^{+'}\end{bmatrix}
\end{equation}
Noting that the weight sets in \cite{mosig1} were computed for \emph{arbitrary} monomial power dependence $k_x^q$, one can cross-multiply the $\tilde{a}^q$ factors in the $\{ \omega_p^{+'}\}$ across the respective rows of \eqref{linsys3} to obtain an analogous system, where now the $\{ \tilde{a}(p)^{(q-n)} \}$ factors in the modified form of \eqref{linsys3} represent (up to a constant) successive $s$ derivatives of $(\xi_1+t_o^+s)^q$ evaluated at $s=0$ \footnote{One does not have the $q$-dependent constants in columns three to $M+1$ of the matrix in \eqref{linsys3}. However, one can include these constants and only affect the unneeded coefficients $\{\tilde{b}^+_1,...,\tilde{b}^+_{M-1}\}$ \cite{linsys}.}. Having now matched our linear system to \cite{mosig1}, the $n$th weight ($n=1,2,...,M+1$) for the tier-($M+1$), complex-plane generalization of the new MWA writes as \cite{mosig1}
\begin{equation}
\label{weights}w_n^{(M+1)} =(-1)^{n+1}\begin{pmatrix} M \\ n-1 \end{pmatrix}\tilde{a}(n)^{M-1-q}/\omega_n^+ \end{equation}
with the expression for our best tail integral estimate given as
\begin{equation}
I^{+(M+1)}_1=\frac{\sum_{n=1}^{M+1}w_n^{(M+1)}I_n^{+'}}{\sum_{n=1}^{M}w_n^{(M+1)}}
\end{equation}
The expressions for the $\{ w_n^{(M+1)} \}$ corresponding to Re$[k_x]<0$ tail integral follows analogously. Furthermore, one expects that with a different choice of $\{ \tilde{\phi}_m(N,s) \}$, this derivation can be repeated to develop complex-plane extensions to other SLST algorithmic members.

 From an analytic standpoint, the derivation of the weights for this formulation is complete. However, despite the analytic form of the new MWA weights shown in \eqref{weights} and \cite{mosig1}, in a finite-precision, numerical implementation this casting leads to arithmetic overflow. This drawback, along with the numerically unstable means to recursively update the weights to procure higher-order weight sets, can be easily remedied as follows:
\begin{enumerate}
\item Starting at some tier-$N$ weight set (e.g. set $N=M+1$), multiply all the weights by $\omega_N^+/\tilde{a}(N)^{N-2-q}$. This ensures that the weights remain bounded for all $n$ and $N$.
\item To subsequently obtain a tier-$(N+1)$ weight set from the tier-$N$ set:
\begin{enumerate}
    \item Set $w_{N+1}^{(N+1)}=(-1)^{N+2}/N$.
    \item For the remaining $N$ weights, set $w_{n}^{(N+1)}=d_{n,N}w_{n}^{(N)}$
\end{enumerate}
\end{enumerate}
where
\begin{equation} d_{n,N}=\frac{e^{-\Delta \xi_x^{+'}(\Delta x\sin{\gamma^+}+\Delta z\cos{\gamma^+})}}{N-n+1}\left(\frac{a(N)}{a(N+1)}\right)^{N-2-q}\frac{a(n)}{a(N+1)}\end{equation}

The second proposed formulation employs the MMA \cite{mich2}, as extended to facilitate adaptive tail integral evaluation \cite{sainath}, in conjunction with our complex extension to the truncation error estimates $\{ \omega_p^{+'},\omega_p^{-'} \}$. The formulae to compute arbitrary-order weight sets is given in \cite{mich2}, while the method to recursively find higher-order weight sets is exhibited in \cite{sainath}. Therefore, the reader is referred to these two references for the elementary details.

Between these two formulations, we opt to implement and show validation results for the second formulation based on the MMA. This is because of the first formulation's poor suitability for an adaptive tail integral evaluation scheme, which in turn is due to increasingly higher-order weight sets being the solutions to increasingly ill-conditioned linear systems. Even though we now have available the analytically recast, numerically stable, closed-form expressions for the first formulation's weights and their update scheme (which obviates any potential algorithmic instability exacerbating computed weight errors), the relative accuracy of the computed weights is still fundamentally capped by the linear system's conditioning\footnote{Indeed, as is well-known, for a condition number $CN$ one can expect to lose $\mathcal{O}(\mathrm{log}_{10}[CN])$ digits of precision in computing the weights \cite{trefethen}.}. To illustrate the ill-conditioning of the weight computation, we show in Figure \ref{CN} below, for four different $\bold{r}-\bold{r}'$ geometries, the two-norm condition number $CN$ \cite{linsys} of \eqref{linsys3} as a function of its rank $M'=M+1$ \footnote{We set $T_0=0.25$m, $q$=0, and $\xi_1=\Delta \xi_x=\cos{\gamma^+}\Delta \xi_x^{+'}=\pi/\Delta x$ for all cases in both figures.}:
\begin{enumerate}
\item $(x-x',z-z')=(1,0)$m
\item $(x-x',z-z')=(1000,0)$m
\item $(x-x',z-z')=(1,10)$m
\item $(x-x',z-z')=(1000,10)$m
\end{enumerate}
To confirm that the system matrix ill-conditioning is not due to the complex-plane generalization of the new MWA, in Figure \ref{CN1} we show the two-norm condition number for $\gamma^+=\pi/4$ while in Figure \ref{CN2} we show, for the same four $\bold{r}-\bold{r}'$ geometries, the conditioning for $\gamma^+=0$ (i.e. as if we performed the standard, real-axis MWA from \cite{mosig1}).
\begin{figure}[H]
\centering
\subfloat[\label{CN1}]{\includegraphics[width=3in]{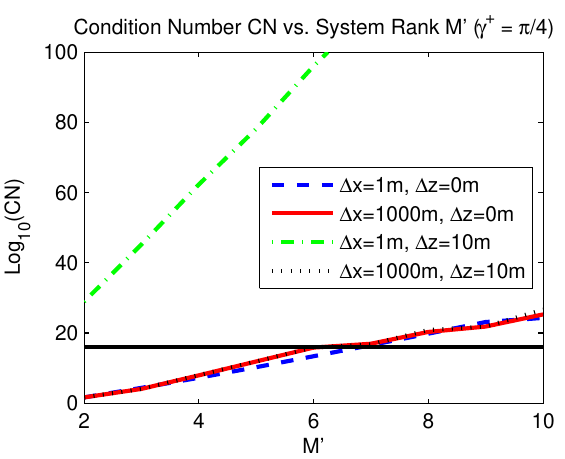}}
\subfloat[\label{CN2}]{\includegraphics[width=3in]{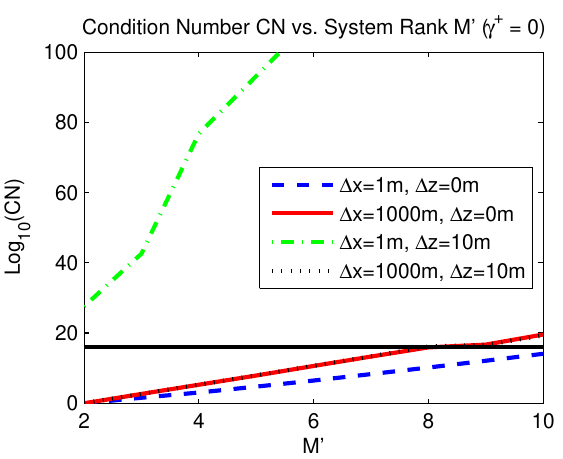}}
\caption{\small Base-10 logarithm of the two-norm system matrix condition number used to compute the new MWA weights, as specified in \eqref{weights} and \cite{mosig1}, for Figures \ref{CN1} and \ref{CN2} (resp.). The vertical axis displays the number of digits of precision lost in the weights, when numerically computing them, due to the conditioning of \eqref{linsys3}. The solid horizontal curve corresponds to Log$_{10}(CN)=16$; weights arising as solutions to a rank-$M'$ linear system with condition number greater than this are expected to be just numerical ``noise" when computed using IEEE double-precision arithmetic.}
\label{CN}
    \end{figure}
One readily observes from Figure \ref{CN} that accurate weight computation is unrealistic as $M'$ increases; in fact, the situation is downright prohibitive for an adaptive MWA implementation (e.g. \cite{sainath}). Even for the best-conditioned geometry (i.e. $(x-x',z-z')=(1,0)$m), one cannot realistically expect even a \emph{single} digit of precision in the weights for $M'$ equalling or exceeding approximately seven and ten in Figures \ref{CN1} and \ref{CN2} (resp.), as can be seen from the intersection of the corresponding curves in Figures \ref{CN1}-\ref{CN2} with the solid horizontal curve corresponding to Log$_{10}(CN)=16$. As a result, we choose the second proposed MWA formulation, based on the MMA \cite{mich2}, for computing validation results in Section \ref{results}. Based on our previous work using the standard, real-axis MMA \cite{sainath} for environments containing high loss and conductively-uniaxial layers, one can expect its success in again producing high-precision results. Indeed, the validation results in Section \ref{results} speak to this effect\footnote{Due to the intricate, recursively-related nature of the MMA weight set tiers, a straightforward definition and analysis of the conditioning of the problem related to procuring these weights proves elusive and therefore is not pursued here. Rather, its use herein is strictly based on, and justified by, its \emph{empirically}-demonstrated efficacy in \cite{mich2,sainath}.}.
\section{\label{results}Results and Discussion}
In this section we exhibit validation results in scenarios involving the modeling of induction sondes for geophysical prospection of hydrocarbons (i.e. induction well logging \cite{zhdanov}). Previously, we demonstrated numerous simulated resistivity logs pertaining to environments containing a combination of isotropic and reciprocal, electrically uniaxial media \cite{felsen} as probed by longitudinally-oriented induction sondes \cite{sainath}. For those case studies, the adaptive, real-axis MMA was successfully incorporated into our algorithm to yield high-precision results exhibiting excellent agreement with data from previous literature \cite{anderson1,wei,howard}.

\begin{figure}[H]
\centering
\subfloat[\label{Sofia4a1R}]{\includegraphics[width=3in]{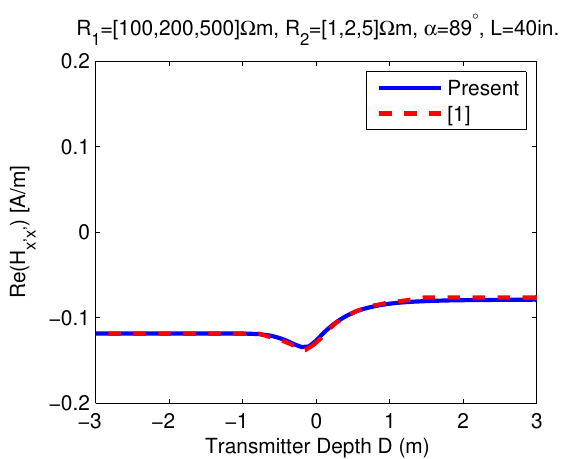}}
\subfloat[\label{Sofia4a1I}]{\includegraphics[width=3in]{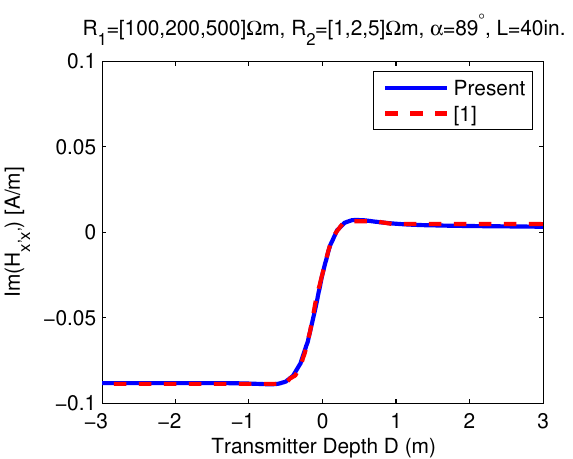}}

\subfloat[\label{Sofia4a2R}]{\includegraphics[width=3in]{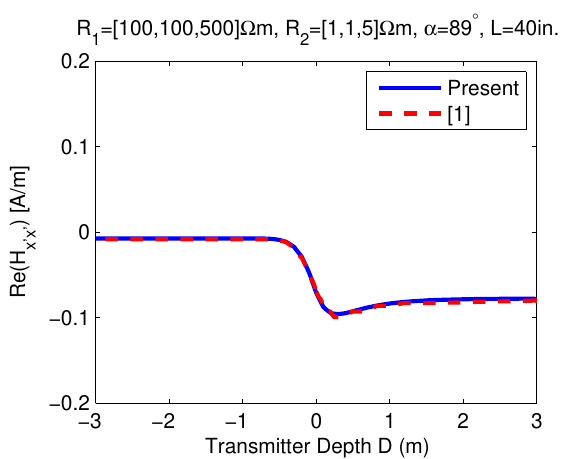}}
\subfloat[\label{Sofia4a2I}]{\includegraphics[width=3in]{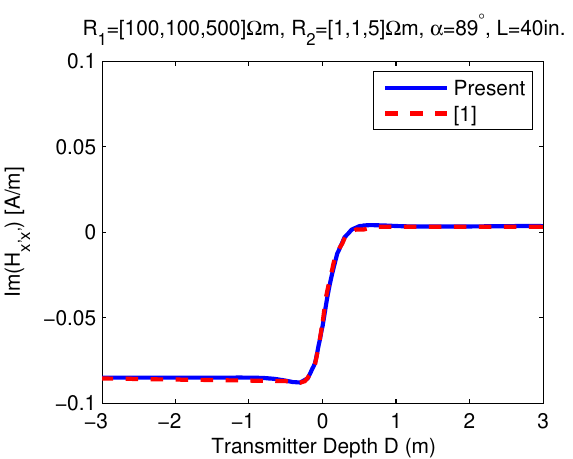}}

\subfloat[\label{Sofia4a3R}]{\includegraphics[width=3in]{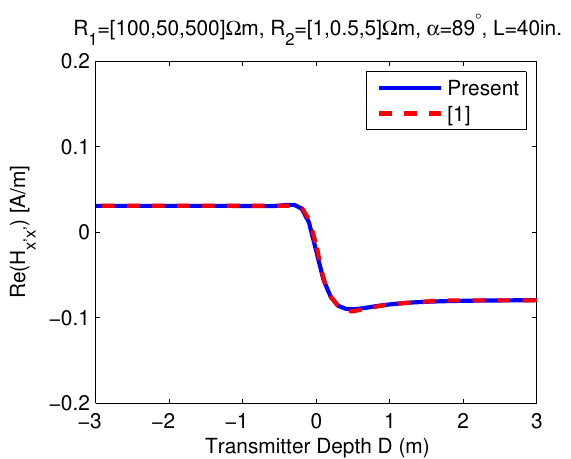}}
\subfloat[\label{Sofia4a3I}]{\includegraphics[width=3in]{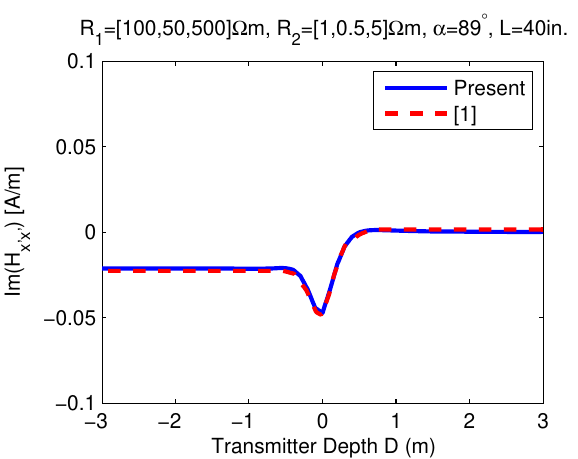}}
\caption{\small Comparison of simulated magnetic field $H_{x'x'}$ with Figure 4 of \cite{sofia}.}
\label{SofiaFig4Hxx}
    \end{figure}

\begin{figure}[H]
\centering
\subfloat[\label{Sofia4b1R}]{\includegraphics[width=3in]{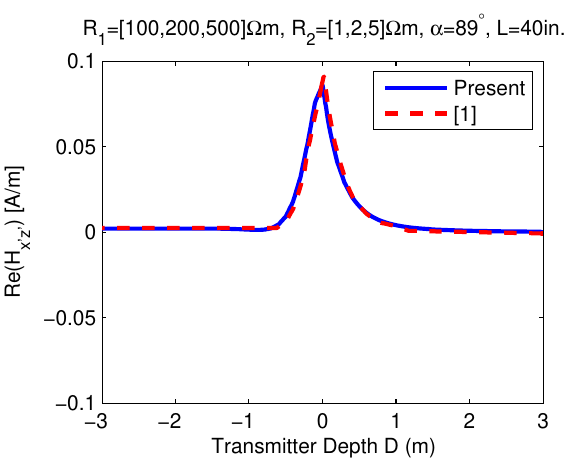}}
\subfloat[\label{Sofia4b1I}]{\includegraphics[width=3in]{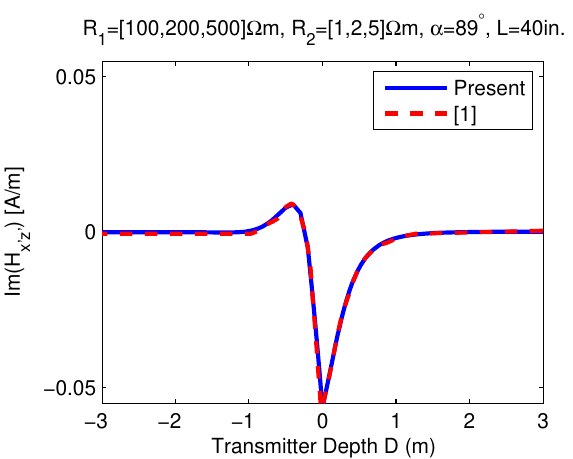}}

\subfloat[\label{Sofia4b2R}]{\includegraphics[width=3in]{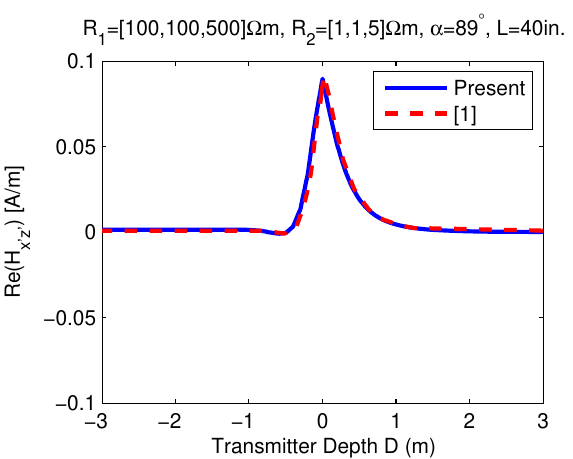}}
\subfloat[\label{Sofia4b2I}]{\includegraphics[width=3in]{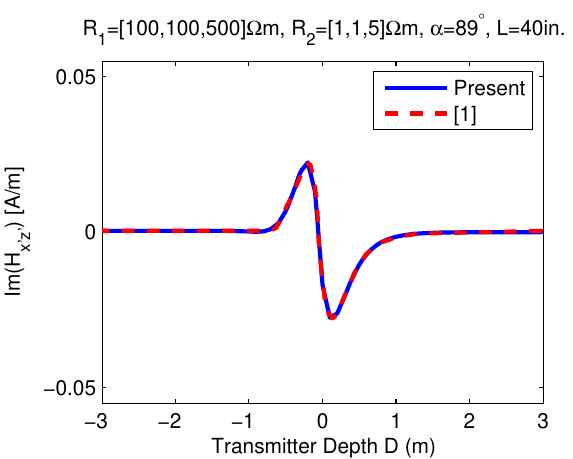}}

\subfloat[\label{Sofia4b3R}]{\includegraphics[width=3in]{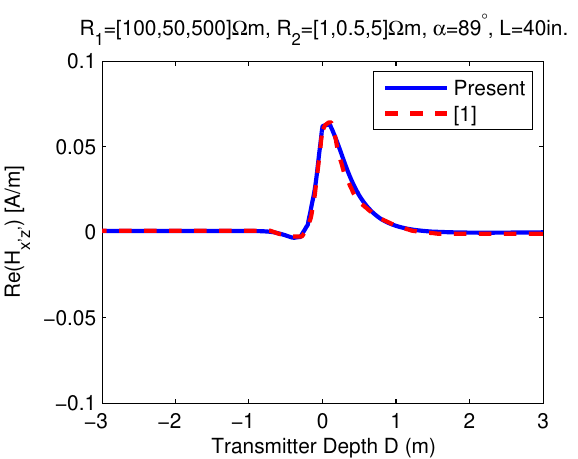}}
\subfloat[\label{Sofia4b3I}]{\includegraphics[width=3in]{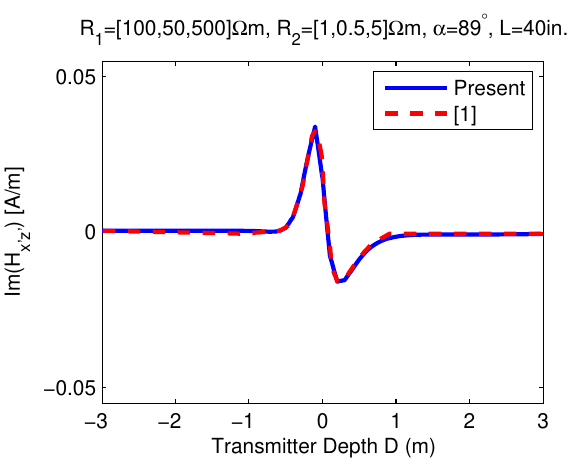}}
\caption{\small Comparison of simulated magnetic field $H_{x'z'}$ with Figure 4 of \cite{sofia}.}
\label{SofiaFig4Hxz}
    \end{figure}

\begin{figure}[H]
\centering
\subfloat[\label{Sofia4d1R}]{\includegraphics[width=3in]{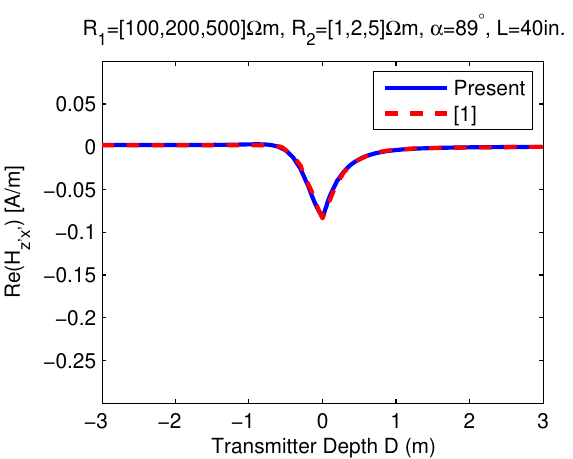}}
\subfloat[\label{Sofia4d1I}]{\includegraphics[width=3in]{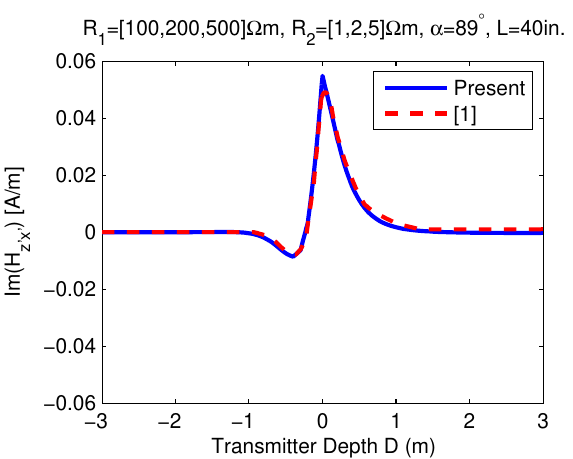}}

\subfloat[\label{Sofia4d2R}]{\includegraphics[width=3in]{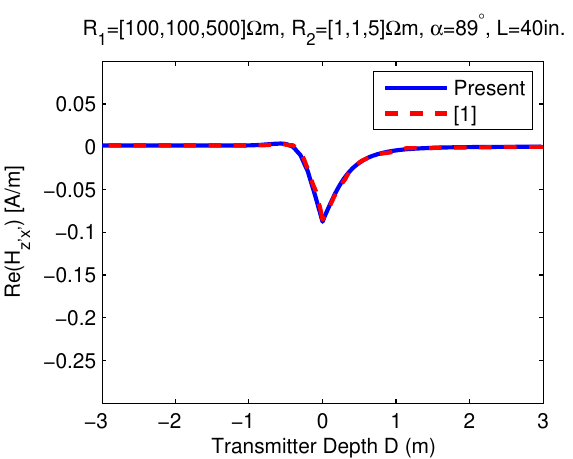}}
\subfloat[\label{Sofia4d2I}]{\includegraphics[width=3in]{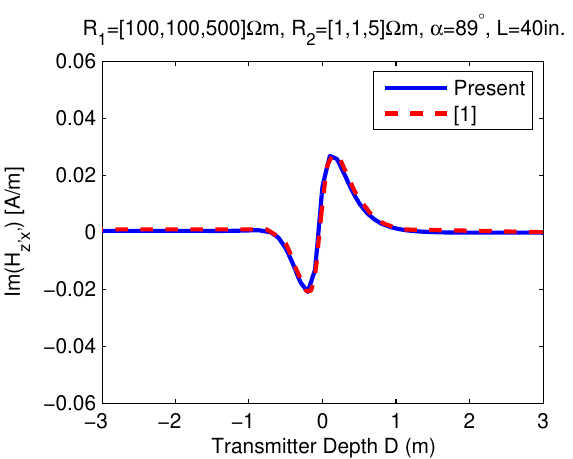}}

\subfloat[\label{Sofia4d3R}]{\includegraphics[width=3in]{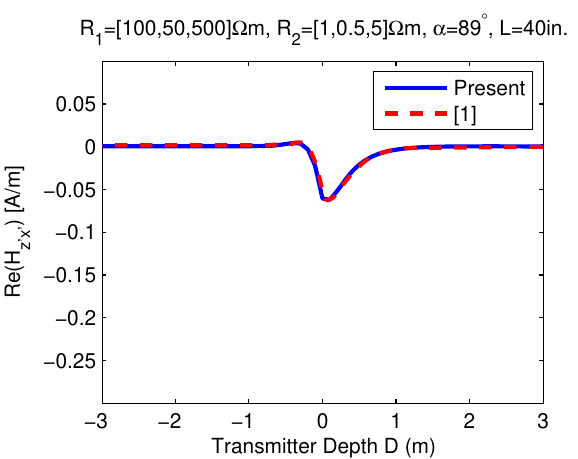}}
\subfloat[\label{Sofia4d3I}]{\includegraphics[width=3in]{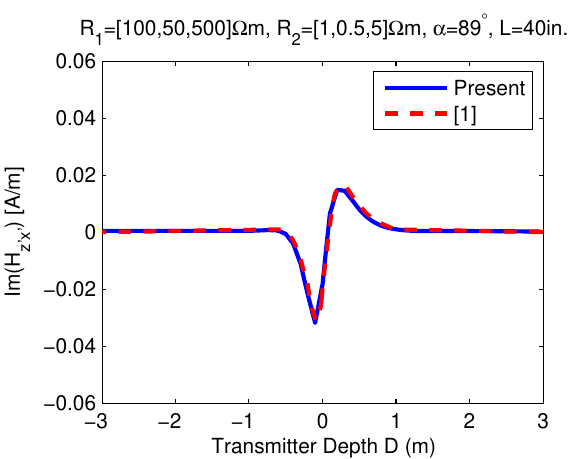}}
\caption{\small Comparison of simulated magnetic field $H_{z'x'}$ with Figure 4 of \cite{sofia}.}
\label{SofiaFig4Hzx}
    \end{figure}

\begin{figure}[H]
\centering
\subfloat[\label{Sofia4e1R}]{\includegraphics[width=3in]{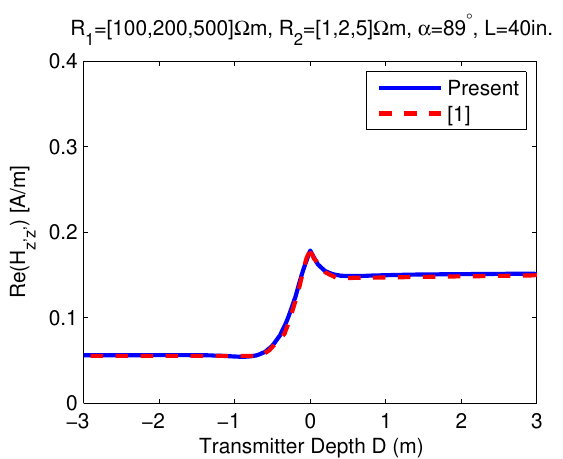}}
\subfloat[\label{Sofia4e1I}]{\includegraphics[width=3in]{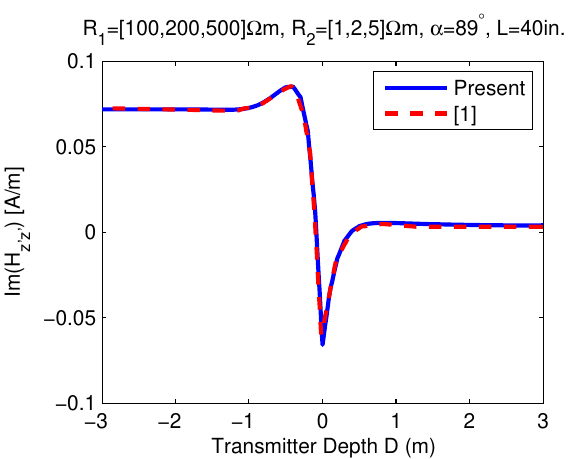}}

\subfloat[\label{Sofia4e2R}]{\includegraphics[width=3in]{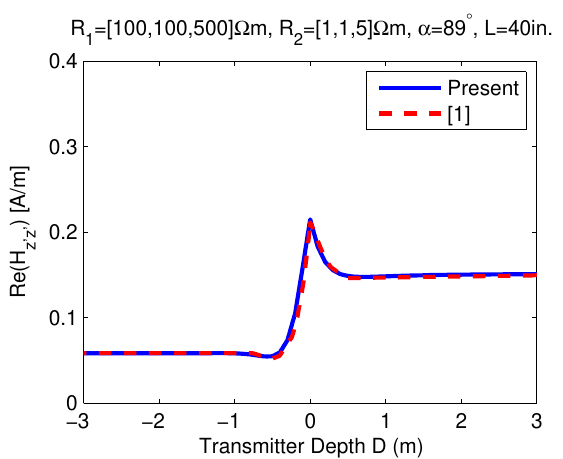}}
\subfloat[\label{Sofia4e2I}]{\includegraphics[width=3in]{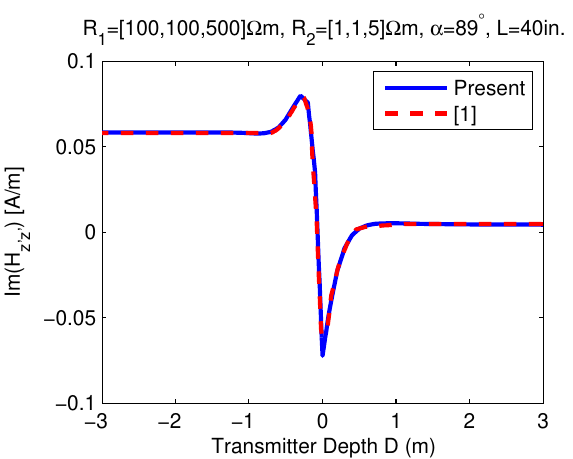}}

\subfloat[\label{Sofia4e3R}]{\includegraphics[width=3in]{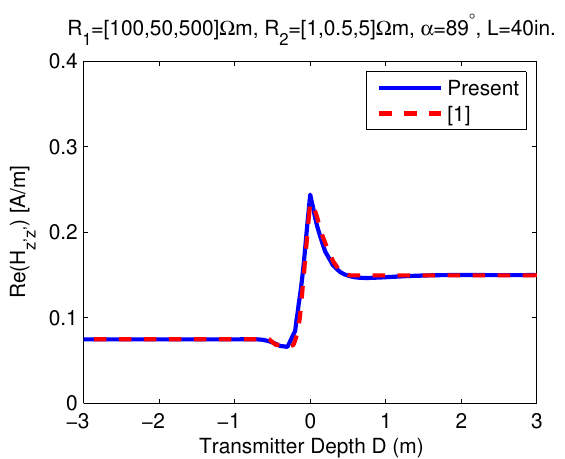}}
\subfloat[\label{Sofia4e3I}]{\includegraphics[width=3in]{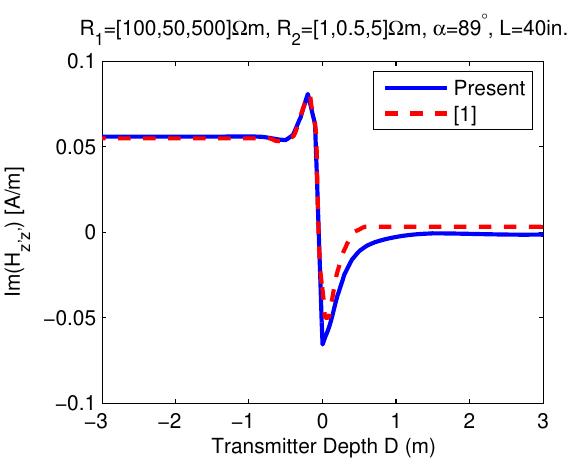}}
\caption{\small Comparison of simulated magnetic field $H_{z'z'}$ with Figure 4 of \cite{sofia}.}
\label{SofiaFig4Hzz}
    \end{figure}

Herein, we exhibit a case study involving a near-horizontal tool orientation where the tool axis dip angle $\alpha=89^{\circ}$, tool axis strike angle $\beta=0^{\circ}$ \footnote{The tool thus rotates and is confined within the $xz$ plane \cite{moran1}.}, and source-observer separation $L_{tool}=|\bold{r}-\bold{r}'|=40"=1.016$m, corresponding to a source-observer depth separation $|z-z'|=L_{tool}\cos{\alpha} \sim 17.7$mm. Consequently, this study serves to validate the efficacy of our new algorithm and its ability to impart absolute, exponential-cum-algebraic convergence on Fourier double-integrals like \eqref{SI1} even for the traditionally prohibitive regime $|z-z'| \ll 1$. Furthermore, to exemplify the general-purpose nature of our new algorithm in regards to the media present, we generate synthetic resistivity logs for a two-layer, planar-stratified environment containing reciprocal, electrically \emph{biaxial} media\footnote{For Figures \ref{SofiaFig4Hxx}-\ref{SofiaFig4Hzz}, the frequency of operation is $f$=2MHz, the interface is located at $z_B=0$m, and the resistivity tensor for layer $n$ (with respect to the standard Earth system) is described in the figure headings by the diagonal matrix $R_n=[R_{x'x'},R_{y'y'},R_{z'z'}]$=diag$[R_{x'x'},R_{y'y'},R_{z'z'}]$.}. In this scenario, wherein all four characteristic plane wave modes in the anisotropic layer containing $\bold{r}$ can (in general) contribute to the observed field, the definition of an exact ``$\Delta z$" and thus CPP is ill-defined (see Section \ref{ext}). Therefore, this set of results also justifies our retaining the MWA's robust environment/source-observer geometry convergence acceleration characteristic, yielding an overall robust and rapid electromagnetic field solution method.

Note that save for Figure \ref{Sofia4e3I}, there exists strong agreement across the full logging path in each plot. Even for Figure \ref{Sofia4e3I}, with some discrepancy in the upper half-space $D>0$, overall there is strong qualitative agreement and (in the bottom half-space) quantitative agreement too\footnote{For the figures shown below, the material formation parameter labeling is reversed versus the labeling in the reference paper such that the material scenario $ \{ R_1=[100,200,500] \Omega \mathrm{m}, \ R_2=[1,2,5] \Omega \mathrm{m} \}$ corresponds to  $ \{ R_1=[100,50,500] \Omega \mathrm{m}, \ R_2=[1,0.5,5] \Omega \mathrm{m} \}$ in \cite{sofia} and vice versa. Due to the strong agreement between the simulation data produced here and by the reference source after this labeling reversal, we suspect this apparent data discrepancy is attributed to a simple typographical error on the part of the authors of \cite{sofia}.}.

\newpage
\section{\label{conv}Convergence Characteristics}
To characterize our numerical formulation's ability to converge towards the field solution, we present two case studies concerning the $z$-directed electric field component $E_z$ produced by a $z$-directed electric dipole radiating at $f$=2MHz in free space. The first case comprises a benign scenario in which $x-x'=y-y'=z-z'=1$m, while the second case represents a very challenging scenario wherein $x-x'=500$m and $y-y'=z-z'=0$m. The latter scenario's prohibitive challenges, when using a standard numerical integration method, are that the integrand (1) oscillates on the order of $500/\sqrt{2}$ times more rapidly than the integrand in case one\footnote{Recall the azimuthal coordinate rotation performed such that in the rotated basis $x-x'=y-y' \geq 0$.} and (2) exhibits absolutely \emph{no} exponential decay due to the annihilation of the exp($i\tilde{k}^+_z|z-z'|$)-type factors. If we were to use a traditional numerical integration methodology, we emphasize that one would obtain a \emph{divergent} result.

For each case, we present results related to the Region III field contribution (see Figure \ref{IntScheme}). Since one cannot obtain a closed-form solution to this field contribution, reference field values from which one measures relative accuracy must be appropriately chosen; their computation details are provided in Figure \ref{ConvFig} below. As in \cite{sainath,mosig1,mich2,mich1}, we assume the integrand is well-behaved in Region III and thus do not perform adaptive interval sub-division. Instead, we set the $k_x$ and $k_y$ plane extrapolation region interval lengths as per Section \ref{ext} and examine the accuracy versus (1) the number of extrapolation region intervals employed ($B$) and (2) the Legendre-Gauss quadrature order used ($LGQ$) to integrate each interval\footnote{$B$ intervals are used in both the $k_x'>0$ and $k_x'<0$ integration path half-tails; the same applies for the $k_y$ path half-tails.}.

Our comments on the relative importance of aliasing and truncation error are analogous to \cite{sainath}: Up to approximately $B=3$ the truncation error dominates the total relative error, while using more than approximately $B=6$ or 7 intervals effects no noticeable decrease in the error for a fixed $LGQ$. Beyond this point aliasing error dominates the total relative error, which is evidenced by the error decreasing versus increasing $LGQ$ but remaining flat versus increasing $B$. However, we notice the following two remarkable characteristics about the algorithm's convergence for case two:
 \begin{enumerate}
 \item The $LGQ=30$ curve reaches within 25dB of case one's $LGQ=30$ curve despite representing a scenario wherein the field solution would ordinarily have \emph{diverged} using standard numerical integration techniques.
 \item Despite this case representing a far more prohibitive scenario (if traditionally evaluated) versus case two presented in \cite{sainath}, wherein $\bold{r}-\bold{r}'=(500,500,1)$m, at $B=10$ the $LGQ=30$ curve here levels off at an error approximately 23dB \emph{lower} than its case two counterpart in \cite{sainath}.
 \end{enumerate}
Note that (akin to, and for the same reasons stated in, \cite{sainath}) relative errors below -150dB were coerced to -150dB.

Since the pre-extrapolation region formulation in this paper is not radically different from that in \cite{sainath}, we expect similar convergence characteristics when using the trapezoidal detour (versus those presented in \cite{sainath}) and thus omit the Region I convergence study for brevity. Furthermore, the Region IIa/IIb convergence studies are omitted as well since the field convergence results would be affected by the algorithm's handling of both the pre-extrapolation and extrapolation region sections of the $k_x$ and $k_y$ plane integration paths. Equivalently, presenting information on the Region I and Region III field convergence characteristics sheds insight into the Region IIa/IIb convergence characteristics. This is because if the respective algorithms handling the Region I and Region III integrations robustly yield accurate, rapidly convergent results, one can expect similar behavior for the Region IIa/IIb results.
\begin{figure}[H]
\centering
\subfloat[\label{Case1Ext}]{\includegraphics[width=3.25in]{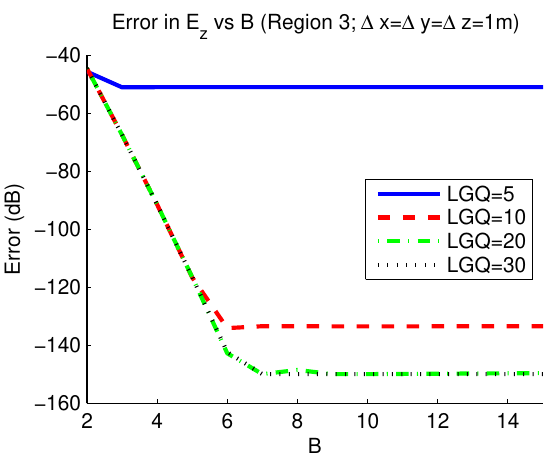}}
\subfloat[\label{Case2Ext}]{\includegraphics[width=3.25in]{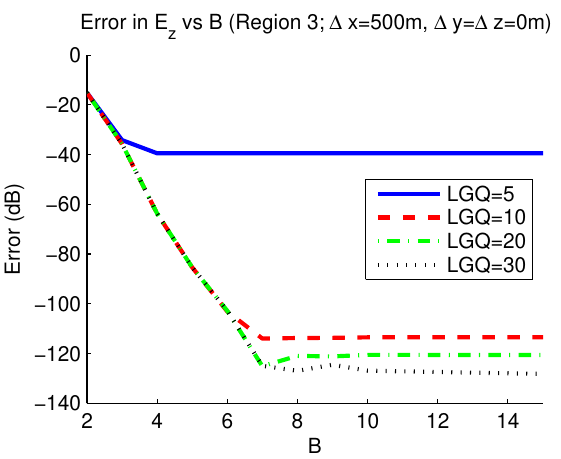}}
\caption{\small Convergence towards the solution comprising the $E_z$ contribution from Region III. The reference field values are computed using $LGQ$=30 and $B=500$ for both figures.}
\label{ConvFig}
    \end{figure}
\section{\label{conclusion}Conclusion}
In this work, we have presented a novel integration scheme composed of (1) a complex-plane, adaptive/error-controlling extension to the standard real-axis MMA in conjunction with (2) a more robust pre-extrapolation region integration path to effect fast, absolute, and \emph{exponential-cum-algebraic} convergence of Fourier- and F-H-type integral transforms such as \eqref{SI1}-\eqref{SI4}. Due to combining the detour with the MMA and its robust algebraic convergence acceleration characteristic, this is indeed the case \emph{irrespective} of the source-observer geometry and loss/anisotropy characteristics of the stratified media present. Furthermore, this is accomplished without the added complication of having to separately account for slab/interface mode contributions whose poles may be crossed when otherwise deforming to more well-known, rapidly-convergent paths such as the SDP \cite{chew2,chewch2}, resulting in a numerically robust and easily-implemented integration methodology.

The algorithm's ability to accurately simulate the observed fields for classically ``worst-case" scenarios $0 \leq |z-z'| \ll 1$, and that too in complex, planar-stratified environments containing biaxial-conductive media, has been verified through numerous validation checks against \cite{sofia}. Finally, the algorithm's convergence characteristics in the strongly-evanescent spectral zone have been explored, analyzed, and shown to be superior compared to an older methodology exhibited in \cite{sainath} that was based on an adaptive extension to the real-axis MMA.

We conclude that the present algorithm's robustness with respect to source-observer geometries and medium types present, as well as its straight-forward nature and ease of implementation, makes it very useful for the analysis of electromagnetic wave propagation and scattering in multi-layered environments containing media of arbitrary anisotropy and loss.
 \section{Acknowledgements}
We thank Halliburton Energy Services for the permission to publish this work. We also acknowledge partial support from the Ohio Supercomputing Center under Grant PAS-0061.
\newpage
\section{References}
\bibliographystyle{model1-num-names}
\bibliography{reflist}

\begin{thebibliography}{37}
\expandafter\ifx\csname natexlab\endcsname\relax\def\natexlab#1{#1}\fi
\providecommand{\url}[1]{\texttt{#1}}
\providecommand{\href}[2]{#2}
\providecommand{\path}[1]{#1}
\providecommand{\DOIprefix}{doi:}
\providecommand{\ArXivprefix}{arXiv:}
\providecommand{\URLprefix}{URL: }
\providecommand{\Pubmedprefix}{pmid:}
\providecommand{\doi}[1]{\href{http://dx.doi.org/#1}{\path{#1}}}
\providecommand{\Pubmed}[1]{\href{pmid:#1}{\path{#1}}}
\providecommand{\bibinfo}[2]{#2}
\ifx\xfnm\relax \def\xfnm[#1]{\unskip,\space#1}\fi
\bibitem[{Davydycheva and Wang(2011)}]{sofia}
\bibinfo{author}{S.~Davydycheva}, \bibinfo{author}{T.~Wang},
\newblock \bibinfo{title}{{M}odeling of {E}lectromagnetic {L}ogs in a
  {L}ayered, {B}iaxially {A}nisotropic {M}edium},
\newblock in: \bibinfo{booktitle}{SEG Annual Meeting}, \bibinfo{year}{2011},
  pp. \bibinfo{pages}{494--498}.
\bibitem[{Wei et~al.(2009)Wei, Wang, and Wang}]{wei}
\bibinfo{author}{B.~Wei}, \bibinfo{author}{T.~Wang}, \bibinfo{author}{Y.~Wang},
\newblock \bibinfo{title}{{C}omputing the {R}esponse of {M}ulti-{C}omponent
  {I}nduction {L}ogging in {L}ayered {A}nisotropic {F}ormation by the
  {R}ecursive {M}atrix {M}ethod with {M}agnetic-{C}urrent-{S}ource {D}yadic
  {G}reen's {F}unction},
\newblock \bibinfo{journal}{Chinese Journal of Geophysics} \bibinfo{volume}{52}
  (\bibinfo{year}{2009}) \bibinfo{pages}{1350--1359}.
\bibitem[{Anderson et~al.(1998)Anderson, Barber, and Gianzero}]{anderson1}
\bibinfo{author}{B.~I. Anderson}, \bibinfo{author}{T.~D. Barber},
  \bibinfo{author}{S.~C. Gianzero},
\newblock \bibinfo{title}{{T}he {E}ffect of {C}rossbedding {A}nisotropy on
  {I}nduction {T}ool {R}esponse},
\newblock in: \bibinfo{booktitle}{SPWLA 39th Annual Logging Symposium},
  \bibinfo{year}{1998}, pp. \bibinfo{pages}{1--14}.
\bibitem[{Howard~Jr.(2000)}]{howard}
\bibinfo{author}{A.~Howard~Jr.},
\newblock \bibinfo{title}{{P}etrophysics of {M}agnetic {D}ipole {F}ields in an
  {A}nisotropic {E}arth},
\newblock \bibinfo{journal}{IEEE Transactions on Antennas and Propagation}
  \bibinfo{volume}{48} (\bibinfo{year}{2000}) \bibinfo{pages}{1376--1383}.
\bibitem[{Zhdanov et~al.(2001)Zhdanov, Kennedy, and Peksen}]{zhdanov}
\bibinfo{author}{M.~Zhdanov}, \bibinfo{author}{W.~Kennedy},
  \bibinfo{author}{E.~Peksen},
\newblock \bibinfo{title}{{F}oundations of {T}ensor {I}nduction
  {W}ell-{L}ogging},
\newblock \bibinfo{journal}{Petrophysics} \bibinfo{volume}{42}
  (\bibinfo{year}{2001}) \bibinfo{pages}{588--610}.
\bibitem[{Wang et~al.(2008)Wang, Tao, Yao, Chen, and Yang}]{wang}
\bibinfo{author}{H.~Wang}, \bibinfo{author}{H.~Tao}, \bibinfo{author}{J.~Yao},
  \bibinfo{author}{G.~Chen}, \bibinfo{author}{S.~Yang},
\newblock \bibinfo{title}{{S}tudy on the {R}esponse of a {M}ulticomponent
  {I}nduction {L}ogging {T}ool in {D}eviated and {L}ayered {A}nisotropic
  {F}ormations by using {N}umerical {M}ode {M}atching {M}ethod},
\newblock \bibinfo{journal}{Chinese Journal of Geophysics} \bibinfo{volume}{51}
  (\bibinfo{year}{2008}) \bibinfo{pages}{1110--1120}.
\bibitem[{Moran and Gianzero(1979)}]{moran1}
\bibinfo{author}{J.~Moran}, \bibinfo{author}{S.~Gianzero},
\newblock \bibinfo{title}{{E}ffects of {F}ormation {A}nisotropy on
  {R}esistivity-{L}ogging {M}easurements},
\newblock \bibinfo{journal}{Geophysics} \bibinfo{volume}{44}
  (\bibinfo{year}{1979}) \bibinfo{pages}{1266--1286}.
\bibitem[{Paulus et~al.(2000)Paulus, Gay-Balmaz, and Martin}]{paulus}
\bibinfo{author}{M.~Paulus}, \bibinfo{author}{P.~Gay-Balmaz},
  \bibinfo{author}{O.~J.~F. Martin},
\newblock \bibinfo{title}{{A}ccurate and {E}fficient {C}omputation of the
  {G}reen's {T}ensor for {S}tratified {M}edia},
\newblock \bibinfo{journal}{Phys. Rev. E} \bibinfo{volume}{62}
  (\bibinfo{year}{2000}) \bibinfo{pages}{5797--5807}.
\bibitem[{Pozar and Sanchez(1988)}]{pozar}
\bibinfo{author}{D.~Pozar}, \bibinfo{author}{V.~Sanchez},
\newblock \bibinfo{title}{{M}agnetic {T}uning of a {M}icrostrip {A}ntenna on a
  {F}errite {S}ubstrate},
\newblock \bibinfo{journal}{Electronics Letters} \bibinfo{volume}{24}
  (\bibinfo{year}{1988}) \bibinfo{pages}{729--731}.
\bibitem[{Pozar(1987)}]{pozar2}
\bibinfo{author}{D.~M. Pozar},
\newblock \bibinfo{title}{{R}adiation and {S}cattering from a {M}icrostrip
  {P}atch on a {U}niaxial {S}ubstrate},
\newblock \bibinfo{journal}{IEEE Transactions on Antennas and Propagation}
  \bibinfo{volume}{35} (\bibinfo{year}{1987}) \bibinfo{pages}{613--621}.
\bibitem[{Jain et~al.(2006)Jain, Lotsberg, and Stamnes}]{jain}
\bibinfo{author}{M.~Jain}, \bibinfo{author}{J.~K. Lotsberg},
  \bibinfo{author}{J.~J. Stamnes},
\newblock \bibinfo{title}{{C}omparisons of {E}xact and {P}araxial {I}ntensities
  of {E}lectromagnetic {W}aves {F}ocused into {U}niaxial {C}rystals},
\newblock \bibinfo{journal}{Journal of Optics A: Pure and Applied Optics}
  \bibinfo{volume}{8} (\bibinfo{year}{2006}) \bibinfo{pages}{709}.
\bibitem[{Jehle et~al.(2009)Jehle, Ruegg, Zuberbuhler, Small, and
  Meier}]{jehle}
\bibinfo{author}{M.~Jehle}, \bibinfo{author}{M.~Ruegg},
  \bibinfo{author}{L.~Zuberbuhler}, \bibinfo{author}{D.~Small},
  \bibinfo{author}{E.~Meier},
\newblock \bibinfo{title}{{M}easurement of {I}onospheric {F}araday {R}otation
  in {S}imulated and {R}eal {S}paceborne {SAR} {D}ata},
\newblock \bibinfo{journal}{IEEE Transactions on Geoscience and Remote Sensing}
  \bibinfo{volume}{47} (\bibinfo{year}{2009}) \bibinfo{pages}{1512--1523}.
\bibitem[{Lambot et~al.(2007)Lambot, Slob, and Vereecken}]{lambot1}
\bibinfo{author}{S.~Lambot}, \bibinfo{author}{E.~Slob},
  \bibinfo{author}{H.~Vereecken},
\newblock \bibinfo{title}{{F}ast {E}valuation of {Z}ero-{O}ffset {G}reen's
  {F}unction for {L}ayered {M}edia with {A}pplication to {G}round-{P}enetrating
  {R}adar},
\newblock \bibinfo{journal}{Geophysical Research Letters} \bibinfo{volume}{34}
  (\bibinfo{year}{2007}).
\bibitem[{Lambot et~al.(2006)Lambot, Slob, Vanclooster, and
  Vereecken}]{lambot2}
\bibinfo{author}{S.~Lambot}, \bibinfo{author}{E.~C. Slob},
  \bibinfo{author}{M.~Vanclooster}, \bibinfo{author}{H.~Vereecken},
\newblock \bibinfo{title}{{C}losed {L}oop {GPR} {D}ata {I}nversion for {S}oil
  {H}ydraulic and {E}lectric {P}roperty {D}etermination},
\newblock \bibinfo{journal}{Geophysical Research Letters} \bibinfo{volume}{33}
  (\bibinfo{year}{2006}).
\bibitem[{Mosig and Gardiol(1982)}]{mosig2}
\bibinfo{author}{J.~R. Mosig}, \bibinfo{author}{F.~E. Gardiol},
\newblock \bibinfo{title}{{A} {D}ynamical {R}adiation {M}odel for {M}icrostrip
  {S}tructures},
\newblock \bibinfo{journal}{Advances in Electronics and Electron Physics}
  \bibinfo{volume}{59} (\bibinfo{year}{1982}) \bibinfo{pages}{139--237}.
\bibitem[{Chew(1990)}]{chewch2}
\bibinfo{author}{W.~C. Chew}, \bibinfo{title}{{W}aves and {F}ields in
  {I}nhomogeneous {M}edia}, \bibinfo{publisher}{Van Nostrand Reinhold},
  \bibinfo{year}{1990}, pp. \bibinfo{pages}{45--160}.
\bibitem[{Michalski(2007)}]{mich1}
\bibinfo{author}{K.~A. Michalski},
\newblock \bibinfo{title}{{A}pplication of the {C}omplex {I}mage {M}ethod to
  {E}lectromagnetic {F}ield {C}omputation in {P}lanar {U}niaxial
  {M}ultilayers},
\newblock in: \bibinfo{booktitle}{Workshop on Integral Techniques for
  Electromagnetics}, \bibinfo{address}{Lausanne, Switzerland},
  \bibinfo{year}{2007}.
\bibitem[{Michalski(1998)}]{mich2}
\bibinfo{author}{K.~A. Michalski},
\newblock \bibinfo{title}{{E}xtrapolation {M}ethods for {S}ommerfeld {I}ntegral
  {T}ails},
\newblock \bibinfo{journal}{IEEE Transactions on Antennas and Propagation}
  \bibinfo{volume}{46} (\bibinfo{year}{1998}) \bibinfo{pages}{1405--1418}.
\bibitem[{Sainath et~al.(2013)Sainath, Teixeira, and Donderici}]{sainath}
\bibinfo{author}{K.~Sainath}, \bibinfo{author}{F.~L. Teixeira},
  \bibinfo{author}{B.~Donderici}, \bibinfo{title}{{R}obust {C}omputation of
  {D}ipole {E}lectromagnetic {F}ields in {A}rbitrarily-{A}nisotropic,
  {P}lanar-{S}tratified {E}nvironments}, \bibinfo{year}{2013}.
  \bibinfo{note}{ArXiv:1309.7952v1}.
\bibitem[{Mosig(2012)}]{mosig1}
\bibinfo{author}{J.~Mosig},
\newblock \bibinfo{title}{{T}he {W}eighted {A}verages {A}lgorithm {R}evisited},
\newblock \bibinfo{journal}{IEEE Transactions on Antennas and Propagation}
  \bibinfo{volume}{60} (\bibinfo{year}{2012}) \bibinfo{pages}{2011--2018}.
\bibitem[{Golubovic et~al.(2012)Golubovic, Polimeridis, and Mosig}]{mosig7}
\bibinfo{author}{R.~Golubovic}, \bibinfo{author}{A.~Polimeridis},
  \bibinfo{author}{J.~Mosig},
\newblock \bibinfo{title}{{E}fficient {A}lgorithms for {C}omputing {S}ommerfeld
  {I}ntegral {T}ails},
\newblock \bibinfo{journal}{IEEE Transactions on Antennas and Propagation}
  \bibinfo{volume}{60} (\bibinfo{year}{2012}) \bibinfo{pages}{2409--2417}.
\bibitem[{Alparslan et~al.(2010)Alparslan, Aksun, and Michalski}]{mich5}
\bibinfo{author}{A.~Alparslan}, \bibinfo{author}{M.~Aksun},
  \bibinfo{author}{K.~Michalski},
\newblock \bibinfo{title}{{C}losed-{F}orm {G}reen's {F}unctions in {P}lanar
  {L}ayered {M}edia for {A}ll {R}anges and {M}aterials},
\newblock \bibinfo{journal}{IEEE Transactions on Microwave Theory and
  Techniques} \bibinfo{volume}{58} (\bibinfo{year}{2010})
  \bibinfo{pages}{602--613}.
\bibitem[{Caboussat and Miers(2010)}]{caboussat}
\bibinfo{author}{A.~Caboussat}, \bibinfo{author}{G.~Miers},
\newblock \bibinfo{title}{{N}umerical {A}pproximation of {E}lectromagnetic
  {S}ignals {A}rising in the {E}valuation of {G}eological {F}ormations},
\newblock \bibinfo{journal}{Computers \& Mathematics with Applications}
  \bibinfo{volume}{59} (\bibinfo{year}{2010}) \bibinfo{pages}{338 -- 351}.
\bibitem[{Homeier(2000)}]{homeier}
\bibinfo{author}{H.~H. Homeier},
\newblock \bibinfo{title}{{S}calar {L}evin-{T}ype {S}equence
  {T}ransformations},
\newblock \bibinfo{journal}{Journal of Computational and Applied Mathematics}
  \bibinfo{volume}{122} (\bibinfo{year}{2000}) \bibinfo{pages}{81 -- 147}.
\bibitem[{Mosig and Melcon(2003)}]{mosig5}
\bibinfo{author}{J.~Mosig}, \bibinfo{author}{A.~Melcon},
\newblock \bibinfo{title}{{G}reen's {F}unctions in {L}ossy {L}ayered {M}edia:
  {I}ntegration along the {I}maginary {A}xis and {A}symptotic {B}ehavior},
\newblock \bibinfo{journal}{IEEE Transactions on Antennas and Propagation}
  \bibinfo{volume}{51} (\bibinfo{year}{2003}) \bibinfo{pages}{3200--3208}.
\bibitem[{Mosig and Gardiol(1983)}]{mosig6}
\bibinfo{author}{J.~Mosig}, \bibinfo{author}{F.~Gardiol},
\newblock \bibinfo{title}{{A}nalytical and {N}umerical {T}echniques in the
  {G}reen's {F}unction {T}reatment of {M}icrostrip {A}ntennas and
  {S}catterers},
\newblock \bibinfo{journal}{IEE Proceedings H: Microwaves, Optics and Antennas}
  \bibinfo{volume}{130} (\bibinfo{year}{1983}) \bibinfo{pages}{175--182}.
\bibitem[{Hu and Chew(1999)}]{chew2}
\bibinfo{author}{B.~Hu}, \bibinfo{author}{W.~Chew},
\newblock \bibinfo{title}{{F}ast {S}teepest {D}escent {P}ath {A}lgorithm for
  {E}lectromagnetic {S}cattering {P}roblems in {L}ayered {M}edium},
\newblock in: \bibinfo{booktitle}{IEEE Antennas and Propagation Society
  International Symposium 1999}, volume~\bibinfo{volume}{2},
  \bibinfo{year}{1999}, pp. \bibinfo{pages}{1182--1185 vol.2}.
  \DOIprefix\doi{10.1109/APS.1999.789524}.
\bibitem[{Chew(1990)}]{chewch3}
\bibinfo{author}{W.~C. Chew}, \bibinfo{title}{{W}aves and {F}ields in
  {I}nhomogeneous {M}edia}, \bibinfo{publisher}{Van Nostrand Reinhold},
  \bibinfo{year}{1990}, pp. \bibinfo{pages}{161--209}.
\bibitem[{Moon et~al.(2013)Moon, Teixeira, and Donderici}]{moon}
\bibinfo{author}{H.~Moon}, \bibinfo{author}{F.~L. Teixeira},
  \bibinfo{author}{B.~Donderici}, \bibinfo{title}{{S}table {P}seudoanalytical
  {C}omputation of {E}lectromagnetic {F}ields from {A}rbitrarily-{O}riented
  {D}ipoles in {C}ylindrically {S}tratified {M}edia}, \bibinfo{year}{2013}.
  \bibinfo{note}{ArXiv:1308.3179v1}.
\bibitem[{Mosig(1989)}]{mosig3}
\bibinfo{author}{J.~R. Mosig},
\newblock \bibinfo{title}{Integral equation technique},
\newblock in: \bibinfo{editor}{T.~Itoh} (Ed.), \bibinfo{booktitle}{{N}umerical
  {T}echniques for {M}icrowave and {M}illimeter {W}ave {P}assive {S}tructures},
  \bibinfo{publisher}{Wiley}, \bibinfo{address}{New York, NY},
  \bibinfo{year}{1989}, pp. \bibinfo{pages}{133--213}.
\bibitem[{Chew(1990)}]{chewch1}
\bibinfo{author}{W.~C. Chew}, \bibinfo{title}{{W}aves and {F}ields in
  {I}nhomogeneous {M}edia}, \bibinfo{publisher}{Van Nostrand Reinhold},
  \bibinfo{year}{1990}, pp. \bibinfo{pages}{1--44}.
\bibitem[{Davies(1988)}]{davies}
\bibinfo{author}{K.~T.~R. Davies},
\newblock \bibinfo{title}{{C}omplex-{P}lane {M}ethods for {E}valuating {H}ighly
  {O}scillatory {I}ntegrals in {N}uclear {P}hysics. ii},
\newblock \bibinfo{journal}{Journal of Physics G: Nuclear Physics}
  \bibinfo{volume}{14} (\bibinfo{year}{1988}) \bibinfo{pages}{973}.
\bibitem[{fub(2013)}]{fubini}
\bibinfo{title}{{F}ubini's {T}heorem},
  \bibinfo{howpublished}{\url{http://en.wikipedia.org/wiki/Fubini's_theorem/}},
  \bibinfo{year}{2013}. \bibinfo{note}{[Online]. Accessed August 25, 2013}.
\bibitem[{Chew(1990)}]{chewch7}
\bibinfo{author}{W.~C. Chew}, \bibinfo{title}{{W}aves and {F}ields in
  {I}nhomogeneous {M}edia}, \bibinfo{publisher}{Van Nostrand Reinhold},
  \bibinfo{year}{1990}, pp. \bibinfo{pages}{375--428}.
\bibitem[{lin(2013)}]{linsys}
\bibinfo{title}{{S}ystem of {L}inear {E}quations},
  \bibinfo{howpublished}{\url{http://en.wikipedia.org/wiki/System_of_linear_equations/}},
  \bibinfo{year}{2013}. \bibinfo{note}{[Online]. Accessed August 11, 2013}.
\bibitem[{Trefethen and Bau(1997)}]{trefethen}
\bibinfo{author}{L.~N. Trefethen}, \bibinfo{author}{D.~Bau},
  \bibinfo{title}{{N}umerical {L}inear {A}lgebra}, \bibinfo{publisher}{SIAM:
  Society for Industrial and Applied Mathematics}, \bibinfo{year}{1997}.
\bibitem[{Felsen and Marcuvitz(1994)}]{felsen}
\bibinfo{author}{L.~B. Felsen}, \bibinfo{author}{N.~Marcuvitz},
  \bibinfo{title}{{R}adiation and {S}cattering of {W}aves}, Electromagnetic
  Waves, \bibinfo{publisher}{IEEE Press}, \bibinfo{address}{Piscataway, NJ},
  \bibinfo{year}{1994}, pp. \bibinfo{pages}{740--820}.

\end{thebibliography}
\newpage
\appendix
\section{Appendix: Definitions and Conventions}
We state the following regarding notation used in this paper:
\begin{enumerate}
\item $i$ is the unit-magnitude imaginary number.
\item The time-harmonic field convention used and suppressed throughout this paper is exp($-i\omega t$), where $\omega$ is the angular frequency at which the source distribution radiates.
\item $c$ is the speed of light in free space.
\item $k_o=\omega/c$ is the characteristic wave number of free space.
\item $\bold{r}=(x,y,z)$ denotes the observer location, while $\bold{r}_t=(x,y)$ denotes the transverse observer location with magnitude $\rho=\sqrt{x^2+y^2}$.
\item $\bold{r}'=(x',y',z')$ denotes the source location, while $\bold{r}_t'=(x',y')$ denotes the transverse source location with magnitude $\rho'=\sqrt{x^{'2}+y^{'2}}$.
\item $\bold{k}=(k_x,k_y,k_z)$ denotes the wave vector, while $\bold{k}_t=(k_x,k_y)$ denotes the transverse wave vector with complex amplitude $k_{\rho}=\sqrt{k_x^2+k_y^2}$. It is implicitly understood that one evaluates $\sqrt{k_x^2+k_y^2}$ such that $H^{(1)}_n(k_{\rho}|\rho-\rho'|)$ (see definition 10 below) corresponds to an exponentially decaying function versus increasing $|\rho-\rho'|$.
\item The axial wave number component for the $n$th characteristic mode supported in layer $M$, $\tilde{k}_{M,nz}$, is coupled to the transverse wave numbers $k_x$ and $k_y$ via the $n$th mode's dispersion relation.
    \item The up-going or down-going mode's axial wave number component in some isotropic layer is denoted by $\tilde{k}_z^+$ or $\tilde{k}_z^-$ (resp.) when the particular layer is not critical to understanding the discussion.
\item $H^{(1)}_n(k_{\rho}|\rho-\rho'|)$ is the order-$n$ Hankel function of the first kind, corresponding to an out-going cylindrical wave.
\item Quantities dependent \emph{only} on one or more spectral variables $\{ k_x,k_y,k_z \}$ are denoted spectral quantities and are distinguished with an over-tilde (e.g. $\tilde{f}(k_x,k_y,k_z)$).
\item Quantities dependent on $\bold{k}_t$, $z$, and $z'$ are denoted mixed-domain quantities and have no over-symbol.
\item Numbers expressed as $\begin{pmatrix} A \\ B \end{pmatrix}$ correspond to the binomial coefficients.
\item $\epsilon_{m,r}$ and $\mu_{m,r}$ represent the relative electric permittivity (including conductive and polarization losses) and relative magnetic permeability of isotropic layer $m$.
\item Re$[k_x]$ and Im$[k_x]$ are used interchangeably with $k_x'$ and $k_x^{''}$ (resp.) to denote the real and imaginary part of $k_x$ (resp.). Analogous definitions apply for $k_y$ and other complex-valued quantities.
    \end{enumerate}
\end{document}